\documentclass[ reprint,superscriptaddress,aps,pre]{revtex4-2}
\usepackage{url}
\usepackage[colorlinks=true,linkcolor=blue,citecolor=blue,urlcolor=blue]{hyperref}

\usepackage{graphicx}%
\usepackage{multirow}%
\usepackage{amsmath,amssymb,amsfonts}%
\usepackage{amsthm}%
\usepackage{mathrsfs}%
\usepackage[title]{appendix}%
\usepackage{xcolor}%
\usepackage{textcomp}%
\usepackage{booktabs}%
\usepackage{algorithm}%
\usepackage{algorithmicx}%
\usepackage{algpseudocode}%
\usepackage{listings}%
\usepackage{upgreek}
\usepackage{mathtools} 
\usepackage{comment} 
\newcommand{\RNum}[1]{\uppercase\expandafter{\romannumeral #1\relax}}

\begin{document}
\title{From Continuous-Time Random Walks to Laplace Tails}

\author{Omer Hamdi}
\affiliation{Physics Department, Bar-Ilan University, Ramat-Gan, 52900, Israel}
\affiliation{Institute of Nanotechnology and Advanced Materials, Bar-Ilan University, Ramat-Gan, 52900, Israel}
\author{Stanislav Burov}
\affiliation{Physics Department, Bar-Ilan University, Ramat-Gan, 52900, Israel}
\author{Eli Barkai}
\affiliation{Physics Department, Bar-Ilan University, Ramat-Gan, 52900, Israel}
\affiliation{Institute of Nanotechnology and Advanced Materials, Bar-Ilan University, Ramat-Gan, 52900, Israel}


\begin{abstract}
During Brownian motion, the displacement is normally distributed, a classical fact aligned with the central limit theorem.
However, single particle tracking in complex media such as glasses, living cells, and colloidal suspensions often reveals pronounced exponential decay of the displacement distribution, known as Laplace tails.
In a short letter, two of us presented the emergence of Laplace tails in the continuous time random walk (CTRW) framework.
Here, a detailed complementary study is presented.
By exploring the behavior of $Q_t(n)$, the probability that exactly $n$ renewals occur during time $t$, we develop a rate function–like framework for this quantity, valid for finite $t$.
We show that $Q_t(n)$ exhibits exponential tails, which in turn give rise to exponential tails of the positional probability density function $P(x,t)$.
Favorable comparison to finite-time numerical simulations and asymptotic large deviation rate functions establishes the validity of our results over a wide temporal range.
\end{abstract}

\maketitle
\section{Introduction}
Brownian motion anchors our understanding of diffusion, with freely diffusing particles exhibiting a normal displacement profile, supported by the central limit theorem (CLT).
However, recent studies have reported the emergence of exponential-like tails in probability density functions (PDFs).
Those tails are referred to as \textit{Laplace tails} \cite{hamdi2024laplace,akimoto2025anomalousstatisticslangevinequation}, due to the statement made by Laplace in 1774 \cite{laplace1774memoire}, that the frequency of an error could be expressed as an exponential of the magnitude of the error, in absolute value.
Laplace had a second law of errors published in $1778$ \cite{laplace1781memoire}, stating that the frequency of the error is an exponential of a quadratic function of the error. This second law is what people call today the Gaussian or the Normal distribution. 

Using single molecule tracking data and computer generated trajectories of a variety of tracers diffusing in disordered media, a growing body of experimental and numerical evidence 
\cite{yuan2024creep,zu2021emergent,kumar2023anomalous,witzel2019heterogeneities,cherstvy2019non,doi:10.1021/acs.jpcb.4c01403,hu2023triggering,doi:10.1021/acs.nanolett.0c01058,Mejía-Monasterio_2020,pnas.2216497120,LAMPO2017532,C0SM00925C,PhysRevE.86.020901,Soares_e_Silva_2014,Weeks2000627,PhysRevLett.103.198103,STYLIANIDOU20142684,10.1063/1.5128743,srinivasan2024breaking,doi:10.1021/acs.jpclett.9b01106,wang2009anomalous,wang2012brownian,rusciano2022fickian,perego2022microscopic,PhysRevResearch.2.022020,guan2014even,PhysRevLett.126.158003,pastore2022model,waigh2023heterogeneous,chaudhuri2007universal,joung2025nanoscale,aaberg2021glass,corci2023extending,miotto2021length,wang2025situ,hidalgo2026telomeres} 
reveals a striking deviation from this classical Gaussian paradigm in many complex systems: the observed probability density of particle displacements, namely $P(x,t)$ often exhibits pronounced exponential-like tails rather than the purely Gaussian decay predicted by the standard Brownian theory. 

Several models, such as diffusing diffusivity models \cite{PhysRevLett.113.098302,yamamoto2021universal,waigh2023heterogeneous,e23020231}, a phenomena known as Brownian yet non-Gaussian diffusion \cite{wang2009anomalous,PhysRevX.7.021002,wang2012brownian,baldovin2019polymerization}, and Fickian yet non-Gaussian diffusion \cite{rusciano2022fickian,PhysRevResearch.2.022020,guan2014even,PhysRevLett.126.158003,pastore2022model} were used to describe this phenomenon.
A different approach was initiated by Chaudhuri, Berthier, and Kob \cite{chaudhuri2007universal}, who analyzed four different systems. Focusing on particle displacements near glass and jamming transitions, highlighting behaviors like sticking (caging) and rapid jumps between basins.
This type of dynamics is described by the continuous time random walk (CTRW) model \cite{hu2023triggering,montroll1965random,BARKAI200213,dybiec2010subordinated,kusmierz2019subdiffusive,aghion2018asymptotic,kutner2017continuous,metzler2000random,Shafir_2022,barkai2000continuous,klafter2011first,Vitali_2022,PhysRevE.109.014130,defaveri2025diffusionquenchedrandomenvironments,burov2022exponential,wang2020large}, which will also be analyzed in this paper. Using specific waiting time distributions, they showed how caging and jumping can be used to predict exponential tails, in accordance with Laplace's first law.
This theory was later advanced by two of the present authors  \cite{barkai2020packets}, showing that it holds under very general settings.
They have shown that, for any distribution of waiting times, which is analytical in the short time limit, and for any jump
length distribution, which decays faster than exponential, Laplace's first law holds at the tails of $P(x,t)$. 
The analysis was based on large deviations \cite{wang2020large,TOUCHETTE20091,PhysRevLett.102.060601,PhysRevLett.113.078101,PhysRevLett.113.120601,PhysRevLett.121.090602,Derrida_2007,du2023dynamical,PhysRevE.103.042116,PhysRevE.109.L022102} arguments, and a saddle point approximation \cite{lugannani1980saddle}, namely to put things in a historical context, a technique used as an extension of the Laplace method for solving integrals \cite{laplace1774memoire}.

Lastly, in our recent paper \cite{hamdi2024laplace}, we analyzed a CTRW with exponential waiting times and symmetric jump distributions. We showed that the presence of universal Laplace tails is determined by the decay rate of the jump-length probability distribution function.
For super-exponential jumps, we showed that the propagator develops Laplace (exponential-like) tails. For sub-exponential jumps, we demonstrated the big jump principle \cite{bassanoni2026bigjumpperturbativeapproach,bassanoni2024rare,singh2023universal}, where the tails are specific to the distribution of the jump distribution.

In the present work, we go beyond this Markovian setting by allowing non-exponential waiting time densities $\psi(\tau)$.
We provide a more detailed analysis of CTRW dynamics, focusing on the waiting times between renewal events $\psi(\tau)$, i.e., the time a particle waits between consecutive jumps. 
In the first part, we focus on $Q_t(n)$, the probability that exactly $n$ renewals occur during time $t$, a crucial ingredient of CTRW.
$Q_t(n)$ captures the temporal structure of any renewal-like process, appearing naturally in diverse areas such as stochastic processes, queuing theory, survival analysis, and a wide range of biological and physical systems. It extends far beyond the context of CTRW.
Using tools from large deviations theory (LDT), we explore the behavior of $Q_t(n)$ as a function of $n/t$.
Specifically, in the large $n/t$ limit, $Q_t(n)$ exhibits an exponential-like tail under very minimal assumptions.
Our theoretical predictions are verified by comparison to exactly solvable examples and numerical trajectory sampling.
The regular focus of LDT is the behavior of $-\log[Q_t(n)]/t$ in the asymptotic $t \to \infty$ limit, neglects all the the exponential pre-factors. Our approach is different: we aim for a better match to experimental data that is obtained for finite $t$.

In the second part of this work we use the $\phi(x|n)$, which denotes the probability of a displacement $x$ after $n$ jumps and condition on the number of renewals obtained by time $t$. This subordination approach, and the previously obtained expression for $Q_t(n)$ allow us to obtain the full form of the propagator $P(x,t)$.
We consider a compressed exponential jump distribution and derive an analytical solution that demonstrates the emergence of exponential-like (Laplace) tails.
Our findings are verified by comparison to numerical simulations of several examples.

The form of $P(x,t)$ developed in this work includes the exponential rate function, and the pre-factors. This allows comparison to experimental and numerical data, obtained at finite measurable time.
In the $t\to \infty$, our results provide the LDT rate function; an object that was explored in the recent work of Pacheco-Pozo and Sokolov \cite{PhysRevE.103.042116} using different methods. 
While our results stem from taking the limit of $t\to\infty$ of $P(x,t)$, obtained for finite $t$, in \cite{PhysRevE.103.042116} the authors operated with LDT rate function expressions of $Q_t(n)$ and $\phi(x|n)$. 
Nevertheless, we demonstrate the asymptotic agreement of the two approaches in the large-$t$ limit.

%
%
%
%
%
%
\section{CTRW Model}
We consider a continuous-time random walk, where a particle (random walker) performs instantaneous jumps separated by random waiting times. The sequence of jump times constitutes a renewal process: the waiting times between successive jumps are independent and identically distributed (IID) random variables  (RVs) drawn from a common probability density function (PDF) $\psi(\tau)$.
Here, $n$ is the number of renewals in the interval $(0,t)$, and $Q_t(n)$ is the probability that exactly $n$ renewals occur during time $t$.

The first jump is at $\tau_1$, the second at $\tau_1 + \tau_2$, and so on, such that $\{\tau_i\}$ are our IID RVs, and $n$ is a random variable.
Many authors have studied $Q_t(n)$ \cite{PhysRevX.4.011028,godreche2001statistics,PhysRevE.108.054113,burov2020limit,cox1962renewal,wang2024statistics,ellettari2025rare,wang2020large}, though one key aspect of this problem seems to have escaped wide recognition; the tail of $Q_t(n)$, i.e., the large $n$ behavior of $Q_t(n)$, exhibits an exponential-like behavior when the waiting time, $\psi(\tau)$, is analytic for small $\tau$ \cite{barkai2020packets}.

The exponential-like tail shows that rare trajectories with unusually large $n$ are not controlled by CLT-type fluctuations around the typical behavior, but instead correspond to paths with many jumps in a given time window, which are only possible when the inter-jump waiting times are short. Therefore, these rare events are sensitive to the short-time form of $\psi(\tau)$, and CLT-based approximations can severely underestimate them.

The final position, $x=\sum_{i=1}^n x_i$, is the sum of $n$ IID continuous jump displacements $x_i$, each drawn from the same time-independent spatial density $f(x_i)$.
%
We denote by $\phi(x|n)$ the PDF of finding the particle at $x$, conditioned on performing exactly $n$ jumps.
Our goal is to determine the PDF $P(x,t)$, i.e., the probability density to find the particle at position $x$ at time $t$, using the subordination relation \cite{metzler2000random,klafter2011first}
\begin{eqnarray}\label{eq_M2:P(x,t)_summation}
P(x,t) = \sum_{n=0} ^\infty Q_t(n) \phi(x|n).
\end{eqnarray}
The sum runs over all possible jump counts $n$ that can occur within the time interval $t$.
We first focus here on $Q_t(n)$, and later, in Section \ref{sec:CTRW_Us}, we analyze $\phi(x | n)$.
Let $\hat{Q}_s(n)=\int_0^\infty \exp(-st)Q_t(n)dt$ be the Laplace transform of $Q_t(n)$, then \cite{feller1991,cox1962renewal,montroll1965random}
\begin{eqnarray}\label{eq_M2:laplacetrans_Q_t(n)}
    \hat{Q}_s(n) = \frac{1-\hat{\psi}(s)}{s}\hat{\psi}^n(s),
\end{eqnarray}
where $\hat{\psi}(s)$ is the Laplace transform of $\psi(\tau)$. 
Equation (\ref{eq_M2:laplacetrans_Q_t(n)}) is easy to derive using $R_n(t)$, the probability density that the $n$-th jump occurred in the interval $(t,t+dt)$, which obeys the renewal relation and yields $Q_t(n)$ via the convolution:
\begin{eqnarray}\label{eq_M2:Q_t(n)_as_func_of_R_t(n)}
    Q_t(n) = \int_0^t W(\tau)R_n(t-\tau)d\tau,
\end{eqnarray}
while $W(t)=1-\int_{0}^{t}\psi(\tau)d\tau$ is the probability that no jump occurred up to time $t$.
Eq. (\ref{eq_M2:Q_t(n)_as_func_of_R_t(n)}) expresses that the $n$-th jump occurs at time $t-\tau$ (with density $R_n(t-\tau)$), and that no further jump occurs in the remaining interval of length $\tau$; integrating over $\tau$ sums over all possible times since the last jump (backward recurrence times). $R_n(t)$ satisfies the equation
\begin{eqnarray}\label{eq_M2:R_t(n)}
    R_n(t) = \int_0^t \psi(\tau)R_{n-1}(t-\tau)d\tau,
\end{eqnarray}
with $R_0(t) = \delta(t)$.
The Laplace transform of Eq. (\ref{eq_M2:Q_t(n)_as_func_of_R_t(n)}), and Eq. (\ref{eq_M2:R_t(n)}) with respect to $t$, and the convolution property leads to Eq. (\ref{eq_M2:laplacetrans_Q_t(n)}).

In Eq. (\ref{eq_M2:laplacetrans_Q_t(n)}), $\hat{\psi}^n(s)$ describes the PDF of the sum of $n$ IID RVs, namely:
\begin{eqnarray}
    \hat{\psi}^n(s) = \left\langle \exp\left[-s(\tau_1+\tau_2+...\tau_n)\right] \right\rangle.\nonumber
\end{eqnarray}
Thus, $\hat{Q}_s(n)$ is related to the classical sum of $n$ IID RVs.
In the special case of exponential waiting times, 
\begin{equation}
    \label{eq:psiExponentialClassical}
    \psi(\tau)=\lambda e^{-\lambda \tau},
\end{equation}
 one has $\hat{\psi}(s)=\lambda/(\lambda+s)$, and Eq. (\ref{eq_M2:laplacetrans_Q_t(n)}) yields the Poisson distribution,
\begin{eqnarray}
    Q_t(n)={\rm e}^{-\lambda t}\frac{(\lambda t)^n}{n!},\nonumber
\end{eqnarray}  
In such case, the typical fluctuations of $n$ become Gaussian at long times, while the decay of $Q_t(n)$ for large $n$ is exponential (according to Stirling approximation for $n!$). 
For general $\psi(\tau)$, $Q_t(n)$ need not be Poisson; nevertheless, when $\langle\tau\rangle$ and $\langle\tau^2\rangle$ are finite, the distribution of $n$ around its mean value is Gaussian.
Throughout the main part of this work, we will assume that $\langle\tau\rangle$ and $\langle\tau^2\rangle$ are finite. See \cite{godreche2001statistics,PhysRevE.98.042139} for the case where these time scales diverge.
In the long-time limit, the mean and variance of $n$ are given by \cite{godreche2001statistics}
\begin{eqnarray}\label{eq_M2:n_moment_1_and_n_moment_2}
    \langle n \rangle &=& \sum_0^\infty n Q_t(n) \underset{t \to \infty}{\approx}  \frac{t}{\langle \tau \rangle},\nonumber\\
    \sigma_n^2 &=& \langle n^2 \rangle -\langle n \rangle^2 \underset{t \to \infty}{\approx}  \frac{\sigma_{\tau}^2}{\langle\tau\rangle^3}t,
\end{eqnarray}
where $\sigma_{\tau}^2=\langle\tau^2\rangle-\langle\tau\rangle^2$ is the variance of the waiting times.
In the long time limit, and when $|n - \langle n \rangle|$ scales as $\sqrt{t}$, corresponding to typical fluctuations around the mean, we have \cite{godreche2001statistics,cox1962renewal,metzler2000random}:
\begin{eqnarray}\label{eq_M2:Q_t_n_Gaussain_general}
    Q_t(n) &\underset{t \to \infty}{\sim}& \exp\left[-\frac{(n-\langle n\rangle )^2}{2 \sigma_n^2}\right]/\sqrt{2 \pi\sigma_n^2}.
\end{eqnarray}
Namely, $Q_t(n)$ is Gaussian, which is expected due to the mentioned connection to the problem of a sum of IID RVs.
While this Gaussian form describes typical fluctuations around the mean, large deviations theory (LDT) offers a global asymptotic description of the distribution, including the tails associated with rare fluctuations. Importantly, as we will show below, the CLT is not replaced but rather embedded within LDT.
%
In the LDT framework, $Q_t(n)$ obeys the large deviation form:
\begin{eqnarray}
\label{eq:qtnlargedeviation}
    Q_t(n) \asymp \exp\left[-t \mathcal{I}\left(\frac{n}{t}\right)\right],
\end{eqnarray}
where $\mathcal{I}\left(y=n/t \right)$ is called the rate function. Here, the considered limit is $n,t \to \infty$, while their ratio $y=n/t$ is finite.
The pre-exponential factors are often important, and their estimation is provided below.
In the following, we show that for any $\psi(\tau)$ that is analytical in the vicinity of $\tau=0$, the rate function in the large $y$ limit is given by
\cite{wang2024statistics}:
\begin{eqnarray}\label{eq_M2:predicted_RF_Structure_Q_T_N}
    \mathcal{I}\left(y = \frac{n}{t}\right) \sim \left(A+1\right)y \ln \left(y\right/c),
\end{eqnarray}
with $c$ a positive constant.
The exponential decay for large $n$, as displayed in Eqs. (\ref{eq:qtnlargedeviation}-\ref{eq_M2:predicted_RF_Structure_Q_T_N}),  implies a nearly exponential decay of $Q_t(n)$ with respect to $n$, which is eventually responsible for the appearance of exponential-like tails in $P(x,t)$ \cite{barkai2020packets}. Here, $A$ is a natural number, related to the waiting time PDF $\psi(\tau)$ as discussed later.
Since our only assumption for $\psi(\tau)$ is the analyticity at short times, the resulting asymptotic exponential decay of $Q_t(n)$ is quite general.
%
\section{Large deviations theory}
We now use standard tools of LDT, exploiting the saddle point approximation technique. For that aim, let $\hat{K}(s) = \ln\left[\hat{\psi}(s)\right]$ be the cumulant generating function of the time intervals between renewals. Using the inverse Laplace transform and Eq. (\ref{eq_M2:laplacetrans_Q_t(n)}), $Q_t(n)$ obtains the form
\begin{eqnarray}
\label{eq_M2:laplacetrans_Q_t(n)_LDT}
    Q_t(n) &=& \frac{1}{2 \pi i}\int_C {\rm e}^{st} \frac{1-\hat{\psi}(s)}{s} \exp\left[n\hat{K}(s)\right]ds\nonumber\\
    &=& \frac{1}{2 \pi i}\int_C  \frac{1-\hat{\psi}(s)}{s} \exp\left[t \left(s+\frac{n}{t}\hat{K}(s)\right)\right]ds.
\end{eqnarray}
The path $C$ is the usual Bromwich contour in the $s$-plane, namely a vertical line in the complex $s$-plane, parallel to the imaginary axis, such that all singularities of 
$\hat{f}(s)$ (e.g., poles or branch cuts) are to the left of the contour.
Using the saddle point method in the framework of LDT \cite{TOUCHETTE20091}, in the limit of large $t$, $Q_t(n)$ attains the form
\begin{eqnarray}\label{eq_M2:Q_t_n_general}
    Q_t(n)&\approx& \frac{1-\hat{\psi}(s_0)}{s_0}\frac{\exp\left[-t\mathcal{I}(y)\right]}{\sqrt{2 \pi |n \hat{K}''(s_0)|}},\\
    \mathcal{I}(y) &=& -s_0-y\hat{K}(s_0).\label{eq_M2:Q_t_n_rate_function_general}
\end{eqnarray}
Here, $\mathcal{I}(y)$ is the rate function, while $s_0$ is the saddle point, deduced from finding the extremum point, namely $\hat{K}'(s_0) = -1/y$.
The pre-factors, $(1-\hat{\psi}(s_0))/s_0\sqrt{2 \pi |n \hat{K}''(s_0)|}$, are crucial for reasonable comparison with data obtained from measurements performed over a finite time $t$.

For a general $\psi(\tau)$, the function $\hat{K}(s)=\ln[\hat{\psi}(s)]$ does not lead to an equation that can be solved explicitly for $s_0$ as a function of $y$. Even when $\hat{\psi}(s)$ is available in a closed form, the saddle-point condition $\hat{K}'(s_0)=-1/y$ is typically a transcendental equation, and therefore a closed form for $s_0$, and consequently for the rate function $\mathcal{I}(y)$, is generally not available.
The rate function $\mathcal{I}(y)$ can nevertheless be obtained for any $y$ in a semi-numerical manner by finding the extremum point $s_0$ using a computer program and plugging it into Eq. (\ref{eq_M2:Q_t_n_rate_function_general}).
We refer to this method as ``numerical large deviation''.
\subsection{Rare events at $y\to\infty$}
Recall that the distribution of the number of jumps, $Q_t(n)$, is approximately Gaussian in the vicinity of $n\sim \langle n\rangle$, as described by Eq. (\ref{eq_M2:Q_t_n_Gaussain_general}). 

When $y=n/t \gg 1$, rare events will occur as the jumps are frequent and we expect large deviations from Gaussianity at finite time $t$.
To accomplish a large number of jumps $n$, one must consider short waiting times between consecutive jumps, controlled by the $\tau \to 0$ limit of $\psi(\tau)$ \cite{barkai2020packets}.
We assume that $\psi(\tau)$ has a Taylor series expansion around $\tau\to 0$, i.e.,
\begin{eqnarray}\label{eq_M2:psi_tau_small_tau_taylor}
    \psi(\tau) \underset{\tau\to 0}{\sim}\sum_{i=0}^\infty C_{A+i}\tau^{A+i},
\end{eqnarray}
where $A$ is an integer representing the first non vanishing term.
Because $\psi(\tau)\ge0$ and $\psi(\tau)\sim C_A \tau^A$ for $\tau \to 0$, $C_A$ must be positive.
Eq. (\ref{eq_M2:psi_tau_small_tau_taylor}), together with a Tauberian theorem \cite{weiss1994aspects}, dictates the corresponding form of the Laplace transform $\hat{\psi}(s)$ in the limit $s \to \infty$, i.e., the large-$s$ regime that probes the short-time structure of $\psi(\tau)$
\begin{eqnarray}\label{eq_M2:psi_s_Taylor}
    \hat{\psi}(s) \underset{s\to \infty}{\sim} \sum_{i=0}^{\infty} \frac{C_{A+i} (A+i)!}{s^{A+i+1}}.
\end{eqnarray}
We now use LDT to compute $Q_t(n)$ in the large $y$ regime. We determine the saddle point $s_0$ by substituting the large $s$ expansion in Eq. (\ref{eq_M2:psi_s_Taylor}) into $\hat{K}'(s)$ and solve the saddle-point condition $\hat K'(s_0)=-1/y$ perturbatively.
Keeping the first two nonzero terms of $\hat\psi(s)$ yields $\hat\psi(s)\sim (C_A A!)s^{-(A+1)}\left[1+(A+1)C_{A+1}/C_As\right]$, and therefore $\hat K'(s)\sim -(A+1)/s - (A+1)C_{A+1}/C_As^2$. A perturbative solution for $\hat K'(s_0)=-1/y$ then gives
\begin{eqnarray}\label{eq_M2:LDT_s_0_large_y}
    s_0 &\approx&(A+1)y+\frac{C_{A+1}}{C_A}.
\end{eqnarray}
Considering additional terms in $\hat\psi(s)$ will yield additional terms for $s_0$ that decay with $y$; the next such term is $( A+2) \left(  2 C_A C_{2 + A}- C_{1 + A}^2 \right)/(A+1) C_A^2
y$ \cite{burov2020limit}, which we do not use in this work.
Substitution of Eq. (\ref{eq_M2:LDT_s_0_large_y}) into Eq. (\ref{eq_M2:Q_t_n_rate_function_general}) yields the LDT large $y$ asymptotical rate function (ARF);
\begin{eqnarray}\label{eq_M2:QtN_RF_APPROX_GENERAL}
    \mathcal{I}(y)\sim  \mathcal{I}_{ARF}(y)=
    (A+1)y\ln\left[\frac{(A+1)y}{e(C_AA!)^{1/(A+1)}}\right]-\frac{C_{A+1}}{C_A}.\nonumber\\
\end{eqnarray}
This implies the nearly exponential decay of $Q_t(n)$ as described in Eq. (\ref{eq_M2:predicted_RF_Structure_Q_T_N}). Higher correction terms can be easily obtained using the next order corrections of $s_0$.
For the final form of Eq. (\ref{eq_M2:Q_t_n_general}), we take only the leading term for the sub-exponential coefficient, i.e., $s_0 \approx (A+1)y,\ \hat{K}''(s_0)\approx (A+1)/s_0^2 $, and $1-\hat{\psi}(s_0) \approx 1$, hence, for large $y$ we obtain:
\begin{eqnarray}\label{eq_M2:Q_t_n_RF_with_pre-factor}
    Q_t(n)&\approx& \frac{1}{\sqrt{(A+1)2\pi n}}\exp\left[-t\mathcal{I}_{ARF}(y)\right].
\end{eqnarray}
Using Stirling's approximation for large $n$, we get a Poisson-like structure for $Q_t(n)$
\begin{eqnarray}\label{eq_M2:Q_T_n_large_y_LDT_Solution}
    Q_t(n) 
    \approx\frac{\left(C_A A!t^{A+1}\right)^n}{(n(A+1))!}\exp\left[\frac{C_{A+1}}{C_A}t\right].
\end{eqnarray}
Thus, for any $\psi(\tau)$ that is analytic for small $\tau$, we get a nearly exponential decay of the distribution of $Q_t(n)$ in the large $n/t$ limit.
This result was obtained previously using combinatorial considerations for any value of $t$ and large $n$ (see in the SM of \cite{barkai2020packets}). 
%
%

While we have just proved the large-$n$ and large-$t$ validity of Eq. (\ref{eq_M2:Q_T_n_large_y_LDT_Solution}) starting from Eq. (\ref{eq_M2:laplacetrans_Q_t(n)_LDT}), the large-$n$, finite-$t$ regime can also be established directly from Eq. (\ref{eq_M2:laplacetrans_Q_t(n)_LDT}) by a saddle point approximation, now treating $t$ as fixed and taking $n\to\infty$. This limit is different from the one used to define the LDT rate function (which is based on $t\to\infty$). Here, we do not re-derive the rate function; instead, we obtain the large-$n$ approximation of $Q_t(n)$ at fixed $t$ and show that it reduces to the same asymptotic form of the LDT approach. Rearranging the terms in Eq.(\ref{eq_M2:laplacetrans_Q_t(n)_LDT}) leads to
\begin{equation}
    Q_t(n) = \frac{1}{2 \pi i}\int_C  \frac{1-\hat{\psi}(s)}{s} \exp\left[n \left(\frac{t}{n}s+\hat{K}(s)\right)\right]ds,
    \label{eq_M2:somewhatunusauleqyation}
\end{equation}
and using the saddle point method in the limit of large $n$ yields
\begin{eqnarray}
    Q_t(n)&\approx& \frac{1-\hat{\psi}(s_0)}{s_0}\frac{\exp\left[-n\left(-\frac{1}{y}s_0-\hat{K}(s_0)\right)\right]}{\sqrt{2 \pi |n \hat{K}''(s_0)|}}.
    %
\end{eqnarray}
Here, $s_0$ is the saddle point, deduced by $\hat{K}'(s_0) = -1/y$. 
We thus obtained an expression equivalent to Eq. (\ref{eq_M2:Q_t_n_general}), derived under the sole assumption that $n$ is large.
Therefore the assumption of large $t$, that turns the sampling of rare events and the rate function into a hard problem, is not necessary. 
It is sufficient to use a finite $t$, and even small $t$, when exploring the large $n$ decay.  

%
%
%
%
%
%
%
%

For the special case when $A=0$, a nice hand waving argument verifies the form of $Q_t(n)$ in Eq. (\ref{eq_M2:Q_T_n_large_y_LDT_Solution}).
To obtain a large $n$, the typical time between jumps is $\tau^\ast = t/n$. Using the short-time expansion
$\psi(\tau) \simeq C_0 + C_1 \tau$, we have
\begin{eqnarray}
\mathrm{Prob}(\tau < \tau^\ast)
\simeq C_0 \tau^\ast + \frac{C_1}{2} (\tau^\ast)^2
= \frac{C_0 t}{n}
\left( 1 + \frac{C_1 t}{2 C_0 n} \right).\nonumber
\end{eqnarray}
We then expect
\begin{eqnarray}
Q_t(n) \simeq \bigl[\mathrm{Prob}(\tau < \tau^\ast)\bigr]^n
\simeq \left( \frac{C_0 t}{n} \right)^n
\left( 1 + \frac{C_1 t}{2 C_0 n} \right)^n .\nonumber
\end{eqnarray}
Using Stirling’s approximation $n! \sim \sqrt{2\pi n},(n/e)^n$ and keeping only the leading terms in the large $n$ limit, this rough estimate gives
\begin{eqnarray}
Q_t(n) \simeq \frac{(C_0 t)^n}{n!}
\exp\left( \frac{C_1 t}{2 C_0} \right).
\end{eqnarray}
Thus, aside from the sub-leading factor $1/2$ appearing in the exponential pre-factor, the estimate reproduces the expression in
Eq. (\ref{eq_M2:Q_T_n_large_y_LDT_Solution}).
\subsection{Rate function at $y\to0$ }\label{sec:small_y_region}
To understand rare events in the regime of small $y = n/t$, one must consider an unusually small number of jumps, $n$, compared to the total time $t$.
The case of $n = 0$ can be solved analytically for a large class of waiting time PDFs given by
 $\psi(\tau) =\sum_{i=1}^{N}c_i \lambda_i\tau^{\alpha_i}{\rm{e}}^{-\lambda_i \tau}$. The probability of not jumping up until time $t$ is:
\begin{eqnarray}\label{eq_M2:Q_t_n=0}
    Q_t(n=0) &=& 1-\int_{0}^t\psi(\tau)d\tau=\int_{t}^{\infty}\psi(\tau)d\tau\nonumber\\
    &=&\sum_{i=1}^{\rm{N}}c_i\lambda_i \int_{t}^{\infty}\tau^{\alpha_i}{\rm{e}}^{-\lambda_i \tau}d\tau.
\end{eqnarray}
In the large $t$ limit, the integral can be approximated by the value of the integrand near the lower bound, since $\psi(\tau)$ has an exponentially decaying tail. Namely,
\begin{eqnarray}
        Q_t(n=0)
        &\approx&\sum_{i=1}^{N}c_i\lambda_i t^{\alpha_i}{\rm{e}}^{-\lambda_i t}.
\end{eqnarray}
The smallest decay rate, $\lambda_{\text{min}}$, determines the dominant term, which yields: $Q_t(n=0) \propto {\rm{e}}^{-\lambda_{{\rm{min}}} t}$. Hence, by definition, the rate function is: 
\begin{eqnarray}\label{eq_M2:rateFuncAty=0_lambda_min}
    \mathcal{I}(y=0)=\lambda_{{\rm{min}}}.
\end{eqnarray}
\subsection{CLT region}
Having characterized the two extreme rare event limits; $y\to\infty$, where large $n$ is realized by atypically short waiting times, and $y= 0$, where no jumps occurred for an exceptionally long time, it remains to connect these regimes through the typical behavior.
In between, the rate function $\mathcal{I}(y)$ interpolates from the tail-dominated values at $y=0$ and $y\to\infty$ to its minimum at the typical jump rate $y_0=1/\langle\tau\rangle$, where the large deviations form reduces to the CLT form and $Q_t(n)$ becomes approximately Gaussian.
Considering the behavior of the rate function in the vicinity of its minimum; that is, expanding the LDT rate function in Eq. (\ref{eq_M2:Q_t_n_rate_function_general}) around its minimum $y_0$, results in 
\begin{eqnarray}\label{eq_M2:CLT_rate_function_LDT_minima}
    \mathcal{I}_{CLT}(y) = \mathcal{I}(y_0)+\frac{1}{2}\frac{d^2}{dy^2}\mathcal{I}(y_0)(y-y_0)^2.
\end{eqnarray}
Our goal now is to identify the typical jump rate $y_0$.
In the typical CLT sector, the dominant saddle point $s_0$ approaches $0$ as $t\to\infty$, so we use the small $s$ expansion of $\hat{K}(s)$. To leading order in $s$, $\hat{K}'(s)\approx-\langle\tau\rangle$, and the saddle point condition $\hat{K}'(s_0)=-1/y_0$ therefore gives the typical value $y_0=1/\langle\tau\rangle$.
Substituting the small $s$ expansion $\hat{K}(s)\approx-\langle\tau\rangle s+\frac{\sigma_\tau^2}{2}s^2$ into Eq. (\ref{eq_M2:Q_t_n_rate_function_general}) and expanding $\mathcal{I}(y)$ around $y_0$ yields $\mathcal{I}(y_0)=0$, and $\mathcal{I}''(y_0)=\langle\tau\rangle^3/\sigma_\tau^2$. Inserting these terms into Eq. (\ref{eq_M2:CLT_rate_function_LDT_minima}) results in
\begin{eqnarray}\label{eq_M2:CLT_Rate_function_general}
    \mathcal{I}(y)\sim\mathcal{I}_{CLT}(y)&=&\frac{(y-1/\langle\tau\rangle)^2}{2\sigma_\tau^2/\langle\tau\rangle^3}.
\end{eqnarray}
Where $\mathcal{I}_{CLT}(y)$ describes the vicinity of the minima of the rate function $\mathcal{I}(y)$ in the limit of $t\to \infty$.

Equivalently, Eq. (\ref{eq_M2:CLT_Rate_function_general}) can be found by inserting Eq. (\ref{eq_M2:n_moment_1_and_n_moment_2}) into Eq. (\ref{eq_M2:Q_t_n_Gaussain_general}), and taking the large $t$ limit. 

%
%
%
\section{Examples for $Q_t(n)$}
In this section, we examine two examples of waiting time distributions \cite{wang2020large} and compare our theoretical predictions with numerical results and simulated sampling, demonstrating large deviations from the CLT predictions when $n\gg t$.
The first is the standard Erlang distribution. The second involves a sum of two exponentials, which can capture distinct short‐time and long‐time scales.
For $t \to \infty$, and $|n - \langle n \rangle| \ll \sqrt{t}$, $Q_t(n)$ of both examples will converge to the CLT.
However, for short $t$, both our examples demonstrate pronounced exponential-like tails.
Even though LDT holds in principle in the limit of large $t$ and $n$, the pre-factor in Eq. (\ref{eq_M2:Q_t_n_RF_with_pre-factor}) plays an important role in obtaining a very good fit to what is measurable in a lab \cite{debiossac2023convergence}. 
\subsection{Erlang distribution}
For the Erlang distribution, an exact expression of $Q_t(n)$ can be compared with our results.
Here,  
\begin{equation}
\psi(\tau) =\frac{\lambda^k \tau^{k-1}}{(k-1)!}\exp{(-\lambda \tau)}, 
\label{eq:earlangpsi01}
\end{equation}
and 
\begin{equation}
Q_{t}(n) =e^{-\lambda t}\sum_{i=nk}^{(n+1)k-1}\frac{(\lambda t)^i}{i!}, 
\label{eq:earlangpsi02}
\end{equation}
 where $k \ge 1$ is an integer shape parameter and $\lambda>0$ is the rate parameter~\cite{forbes2011statistical}.

First, the rare-event regime $y \to \infty$ is analyzed by series expansion of $\psi(\tau)$, which yields $A=k-1,C_A=\lambda^k/(k-1)!,C_{A+1}=-\lambda^{k+1}/(k-1)!$. These coefficients allow us to obtain the asymptotic rate function from Eq. (\ref{eq_M2:QtN_RF_APPROX_GENERAL}) \cite{wang2020large}:
\begin{eqnarray}\label{eq_M2:_RF_Erlang_Large_t}
    \mathcal{I}_{ARF}(y) = yk\ln\left[\frac{k}{\lambda e}y\right] +\lambda .
\end{eqnarray}
The same result can be deduced from the leading term of $Q_t(n)\sim e^{-\lambda t}\frac{(\lambda t)^{nk}}{(nk)!}$, by taking the log and dividing by $-t$.
%

%
Second, for $y \to 0$, we approximate the rate function by analyzing $Q_t(n)$ for small $n/t$. This yields exactly Eq. (\ref{eq_M2:_RF_Erlang_Large_t}). In this limit of $y\to0$, the rate function evaluates to $\lambda$, which is the decay rate of the PDF.
Hence $\mathcal{I}(0)=\lambda$, in agreement with Eq. (\ref{eq_M2:rateFuncAty=0_lambda_min}).
Moreover, we can use the exact expression of $Q_t(n=0)$, i.e.,
\begin{eqnarray}\label{eq_M2:RF_Erlang_n=0}
    Q_t(n=0) = \frac{\Gamma(k, \lambda t)}{(k-1)!} = {\rm{e}}^{-\lambda t}\sum_{m=0}^{k-1}\frac{(\lambda t)^m}{m!},
\end{eqnarray}
yielding the rate function $\mathcal{I}(y=0)=\lambda$, again matching with our prediction.

Third, for the CLT regime, we can either derive the rate function by inserting Eq. (\ref{eq_M2:_RF_Erlang_Large_t}) into Eq. (\ref{eq_M2:CLT_rate_function_LDT_minima}):
\begin{eqnarray}\label{eq_M2:RF_Erlang_CLT}
    \mathcal{I}_{CLT}(y)   = \frac{\left(y-\frac{\lambda}{k}\right)^2}{2\lambda / k^2},
\end{eqnarray}
which by adding the normalization prefactor leads to 
\begin{equation}
Q_t^{CLT}(n) \approx \frac{1}{\sqrt{2\pi\lambda t/k^2}}\exp\left[-t\mathcal{I}_{CLT}(\frac{n}{t})\right].
    \label{eq:qtn_cltprefactor}
\end{equation}
Equivalently, one can derive the first two moments of the Erlang distribution and insert them into Eq. (\ref{eq_M2:CLT_Rate_function_general}) to obtain $\mathcal{I}_{CLT}(y)$.

\begin{figure}[ht]
\centering
\includegraphics[width=0.485\textwidth]{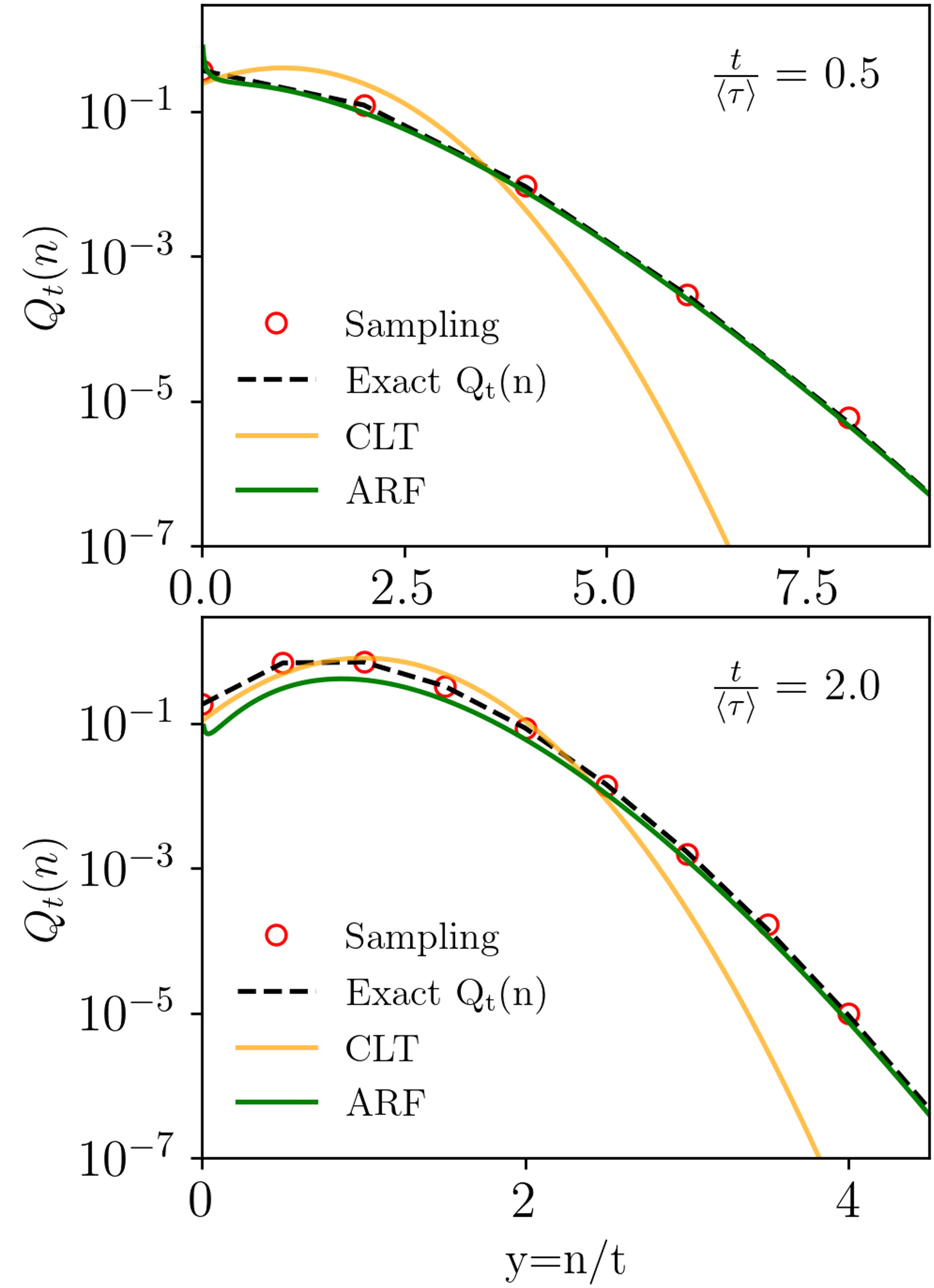}
\caption{
Comparative analysis of $Q_t(n)$ for numerically sampled data using the Erlang distribution (red circles). The exact solution is detailed in the text (black dashed line), CLT predictions (orange line) as provided by Eq.~\eqref{eq:qtn_cltprefactor}, 
and our ARF approximation, produced by inserting Eq. (\ref{eq_M2:_RF_Erlang_Large_t}) into Eq. (\ref{eq_M2:Q_t_n_RF_with_pre-factor}) (green line).
%
Plotted are two example cases with Erlang parameters $k=2,\lambda=2$, and $10^6  $ samples on a semi-log scale. Here, $t/\langle\tau\rangle=0.5$ on top, and $t/\langle\tau\rangle=2$ on the bottom.
As we increase $t$, $Q_t(n)$ will converge to the normal distribution. However, for small values of $t$, we see a good agreement with the ARF theory.
%
}\label{Fig_Qtn_Erlang_compare}
\end{figure}

The question remains: How well do these results fit the finite-time numerical simulations. In Fig. \ref{Fig_Qtn_Erlang_compare}, we compare the form of $Q_t(n)$ produced in this section for the Erlang distribution for two different values of $t$, plotting the exact analytical solution described at the beginning of this section, numerical sampling of the waiting times, CLT predictions depicted in Eq. (\ref{eq:qtn_cltprefactor}), 
and our ARF approximation of $Q_t(n)$, produced by inserting Eq. (\ref{eq_M2:_RF_Erlang_Large_t}) into Eq. (\ref{eq_M2:Q_t_n_RF_with_pre-factor}).
The comparison is performed for finite and practical values of $t$, as opposed to the common practice when the observable $-\log[Q_t(n)]/t$ is compared to  the LDT result in the $t\to \infty$ limit. 

\subsection{Sum of Two Exponential Distributions}
For the second example, we use a sum of two exponentials with different rates \(\lambda_{1}\) and \(\lambda_{2}\), namely
\begin{equation}
    \label{eq:sec4Bpsiexmpl}
    \psi(\tau) = d_0\left(d_1 {\rm{e}}^{-\lambda_1 \tau}+ {\rm{e}}^{-\lambda_2 \tau}\right),
\end{equation}
such that $d_0 =1/\left(\frac{d_1}{\lambda_1}+\frac{1}{\lambda_2}\right)$.
In Appendix \ref{ap:2Exp_Exact_sol}, we outline how we obtained an exact analytical solution for $Q_t(n)$, which we will refer to in this section.
As before, the rate function exhibits the three general behaviors discussed in the previous section.
First, from series expansion of $\psi(\tau)$, we get $A=0,C_A=d_0(d_1+1),C_{A+1}=-d_0(d_1\lambda_1+\lambda_2)$, hence the asymptotical rate function in Eq. (\ref{eq_M2:QtN_RF_APPROX_GENERAL}) is
\begin{eqnarray}\label{eq_M2:_RF_2Exp_Large_t}
        \mathcal{I}_{ARF}(y)     &=&y\ln\left[\frac{1}{ed_0(d_1+1)}y\right]
         +\frac{d_1\lambda_1+\lambda_2}{d_1+1},
\end{eqnarray}
which is inserted into Eq. (\ref{eq_M2:Q_t_n_RF_with_pre-factor}) to produce our ARF approximation for $Q_t(n)$

Second, for $y\to 0$, the exact analytical solution of $Q_t(n)$ at $n=0$ is
\begin{eqnarray}
\frac{\lambda_1 e^{-\lambda_2 t}+\lambda_2 d_1 e^{-\lambda_1t}}{\lambda_1+d_1 \lambda_2}.   
\end{eqnarray}
Hence, according to Eq. (\ref{eq_M2:rateFuncAty=0_lambda_min}) the rate function is exactly $\lambda_{min}$, in the large $t$ limit.

Third, for the CLT regime, we take the first moment and the standard deviation of $\psi(\tau)$; $\langle\tau\rangle = d_0\left(d_1/\lambda_1^2+1/\lambda_2^2\right),\ \sigma_{\tau}^2=d_0 \left( 2 \left( d_1 / \lambda_1^3 + 1 / \lambda_2^3 \right) - d_0 \left( d_1 / \lambda_1^2 + 1 / \lambda_2^2 \right)^2 \right)
$, and substitute these expressions into Eq. (\ref{eq_M2:CLT_Rate_function_general}).

\begin{figure}[ht]
\centering
\includegraphics[width=0.485\textwidth]{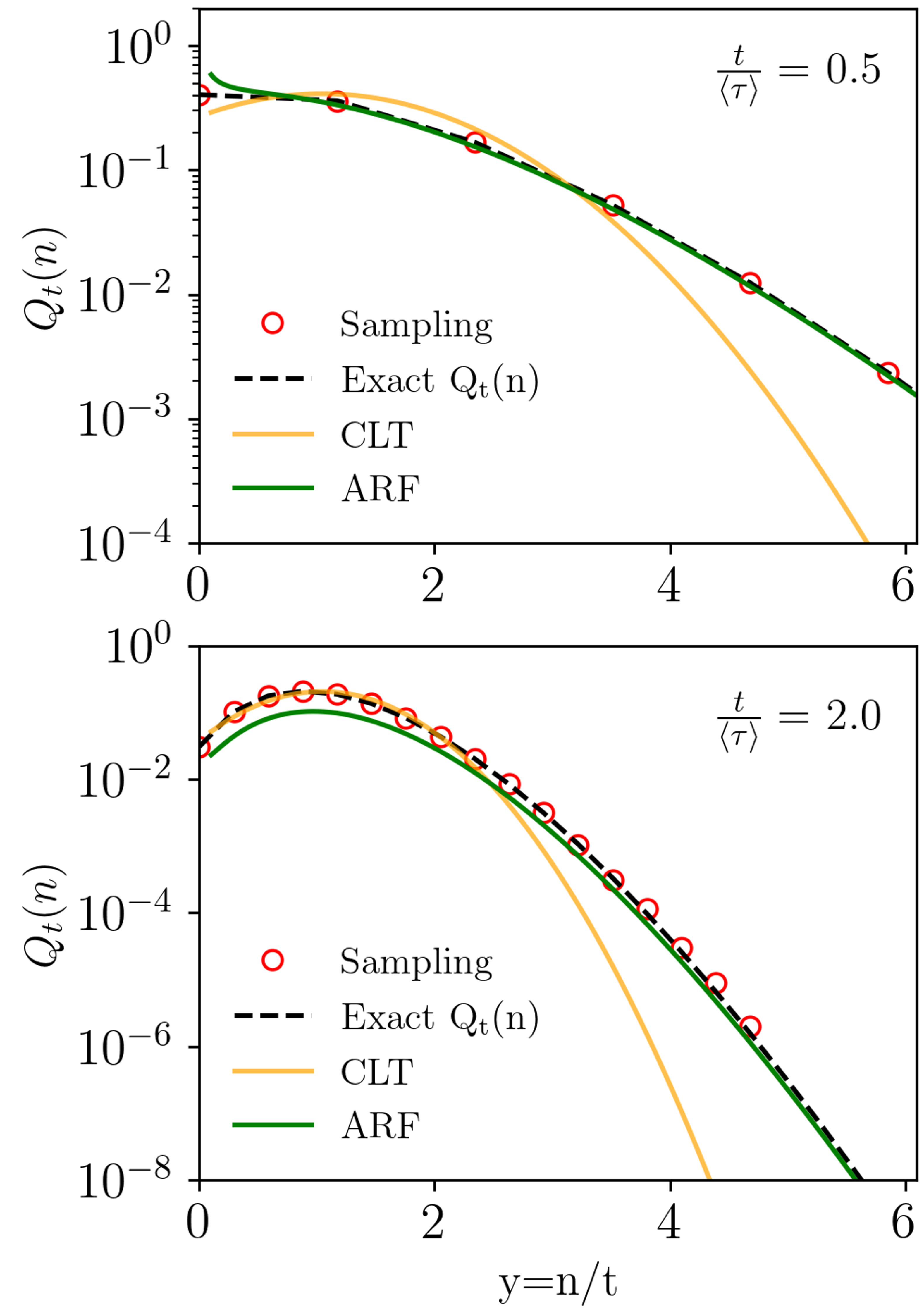}
\caption{
Comparative analysis of $Q_t(n)$ as obtained for $\psi(\tau)$ in Eq.~\eqref{eq:sec4Bpsiexmpl}. The presented behavior is for numerically sampled data (red circles), exact solution from Eq. (\ref{eq_M2:Q_t_n_SumOfTwoExp})
(black dashed line), CLT predictions (orange line),
and our ARF approximation of $Q_t(n)$, produced by inserting Eq. (\ref{eq_M2:_RF_2Exp_Large_t}) into Eq. (\ref{eq_M2:Q_t_n_RF_with_pre-factor}) (green line).
%
Plotted are two example cases with the parameters $d_1=5,\lambda_1=1,\lambda_2=3$, and $10^6$ samples.
Here, $t/\langle\tau\rangle=0.5$ on top, and $t/\langle\tau\rangle=2$ on the bottom.
As $t$ increases, the empirical sampling converges to the CLT. For small $t$, we observe good agreement with the ARF theory.
}\label{Fig_2Exp_Qtn_compare}
\end{figure}

In Fig. \ref{Fig_2Exp_Qtn_compare}, we present the comparison of ARF and CLT approximations produced in this section to exact behavior of $Q_t(n)$ for two distinct values of $t$. 
While for $t\langle /\tau\rangle=2$ the CLT approximation works very well for small values of $y$, the ARF approximation works extremely well for large values of $y$ for both examined cases of $t\langle /\tau\rangle$. 

%
%
%
%
%
\section{CTRW in the large $x$ limit}\label{sec:CTRW_Us}
In this section, we study the behavior of $P(x,t)$ for large $x$, and later obtain the rate function for $P(x,t)$.
The propagator $P(x,t)$ is our primary observable: it is readily measured 
and provides the single-time statistics of the process.
In this section, we use our previously obtained results and the CTRW framework to describe the propagator in the tails, presenting concrete and relevant examples and comparing our method to the CLT predictions.
We limit ourselves to a symmetric jump distribution $f(x_i)$, such that its variance is set to unity, and its decay is faster than exponential, that is, $\underset{x_i\to\infty}{\lim}f(x_i){\rm{e}}^{cx_i} \to 0$ for any $c>0$. 
When the decay is slower than an exponential, we get another phenomenon called the big jump principle \cite{bassanoni2026bigjumpperturbativeapproach,bassanoni2024rare,singh2023universal}, which is out of the scope of this work.

Typically, one approximates $P(x,t)$ by the Central Limit Theorem:
\begin{eqnarray}
\label{eq:pxtCLTFirst}
    P(x,t)_{CLT}=\exp\left[-x^2/4Dt\right]/\sqrt{4\pi Dt},
\end{eqnarray}
$D$ being the diffusion coefficient \cite{einstein1905motion}. In our examples, $D=1/2\langle\tau\rangle$, where we used the fact that the variance of jump lengths is unity.

We focus on the tails of $P(x,t)$, namely the decay in the large $x$ limit.
This is achieved by computing $\phi(x|n)$ and substituting it into Eq. (\ref{eq_M2:P(x,t)_summation}). We begin by noting that the jumps are IID random variables, so $\phi(x | n)$ can be written as the $n$-fold convolution of the single jump PDF $f(x_i)$. In Fourier space \cite{bochner1949fourier}, these convolutions become products:    
\begin{equation}\label{eq_M2:tilde{f}^n}
\widetilde{\phi}^n(k|n)=\int_{-\infty} ^\infty {\rm e}^{ i k x} \phi(x|n) {\rm d} x =
\widetilde{f}^n(k),
\end{equation}
while $ \widetilde{f}(k) = \int_{-\infty} ^\infty \exp( i k x_i) f(x_i) {\rm d} x_i$ is the Fourier transform of a single jump PDF. 
The Fourier transform of $P(x,t)$, i.e., $\widetilde{P}(k,t)$, is obtained from
Eq. (\ref{eq_M2:P(x,t)_summation}) and Eq. (\ref{eq_M2:tilde{f}^n})
\begin{equation}\label{eq_M2:P(k,t)_exact}
\widetilde{P} (k, t)= \sum_{n=0} ^\infty Q_t(n) \widetilde{f}^n(k).
\end{equation}
The goal is to find the inverse Fourier transform of the expression above in the limit of large $x$ and finite $t$.
In this limit of large $|x|$, the magnitude of $f(|x_i|)$ is extremely small and therefore the non-zero contribution to $\phi(x|n)$ must come from a sufficiently large number of steps $n$ \cite{wang2020large}.
Therefore, by fixing $t$ and working in the large $|x|$ limit, the dominant contribution in Eq (\ref{eq_M2:P(k,t)_exact}) comes from the leading large $n$ term of $Q_t(n)$ in Eq. (\ref{eq_M2:Q_T_n_large_y_LDT_Solution}).
In Appendix \ref{ap:from_exact_p(k,t)_to_large)_n_limit} we exploit this property and show that the small $k$ limit, which corresponds to large $x$, results in
\begin{eqnarray}\label{eq_M2:P(k,t)_large_n_limit}
\widetilde{P} (k, t) &\approx&  \exp\left[ \frac{C_{A+1}}{C_A}t+\left[C_A A!\widetilde{f}(k)\right]^{1/(A+1)} t\right].
\end{eqnarray}
The inverse Fourier transform of $\widetilde{P} (k, t)$, while changing variable $i k =u$, reads
\begin{eqnarray}\label{eq_M2:P(x,t)_InvFourier}
P(x,t) &\approx& \frac{1}{2 \pi i} \int_{- i \infty} ^{i \infty} \exp{\left[ - x T(u) \right]}  {\rm d} u,
\end{eqnarray}
where
\begin{eqnarray}\label{eq_M2:KofU_CTRW}
T(u) &=& u - \frac{1}{q} \left[ \frac{C_{A+1}}{C_A}+\left[C_A A!\hat{f}(u)\right]^{1/(A+1)} \right],
\end{eqnarray}
$q \equiv x/t$, and $\hat{f}(u) = \int_{-\infty} ^\infty \exp( u x_i) f(x_i) {\rm d} x_i$ is the moment generating function.
The constraint of $f(x_i)$ decaying faster than an exponential for large $|x_i|$ is prominent here; otherwise, the integral defining $\hat{f}(u)$ diverges, and this method will break.
The problem of finding $P(x,t)$ is solved by using the saddle point approximation: First, find the $u$ that satisfies $T'(u)=0$ and term this $u$ as $u_0$.
Then, expand $T(u)$ in Eq. (\ref{eq_M2:P(x,t)_InvFourier}), in the vicinity of $u_0$, up to a quadratic term. The obtained Gaussian integral yields:
\begin{equation}\label{eq_M2:P(x,t)_LDT_K(u_0)}
P(x,t) \approx \frac{\exp\left[ - x T(u_0)\right]}{ \sqrt{ 2 \pi |x T''(u_0)|} }.
\end{equation}
Since $u_0$ satisfies $T'(u_0)=0$, we obtain
\begin{equation}\label{eq_M2:exact_u_0_forumula}
q = \left(C_A A!\right)^{\frac{1}{A+1}} \hat{f}(u_0)^{\frac{1}{A+1}-1}\hat{f}'(u_0).
\end{equation}
From this equation, we expect $u_0$ to grow as a function of $q=x/t$ and large $q$ corresponds to large $u_0$. 
We will show that in this limit of large $q=x/t$ and large $u_0$ the growth of $T(u_0)$ is very slow, that   ,according to Eq.~\eqref{eq_M2:P(x,t)_LDT_K(u_0)}, leads to appearance of Laplace tails.
In the next section, we test the theory derived above using specific examples.

\begin{figure}[ht]
\centering
\includegraphics[width=0.486\textwidth]{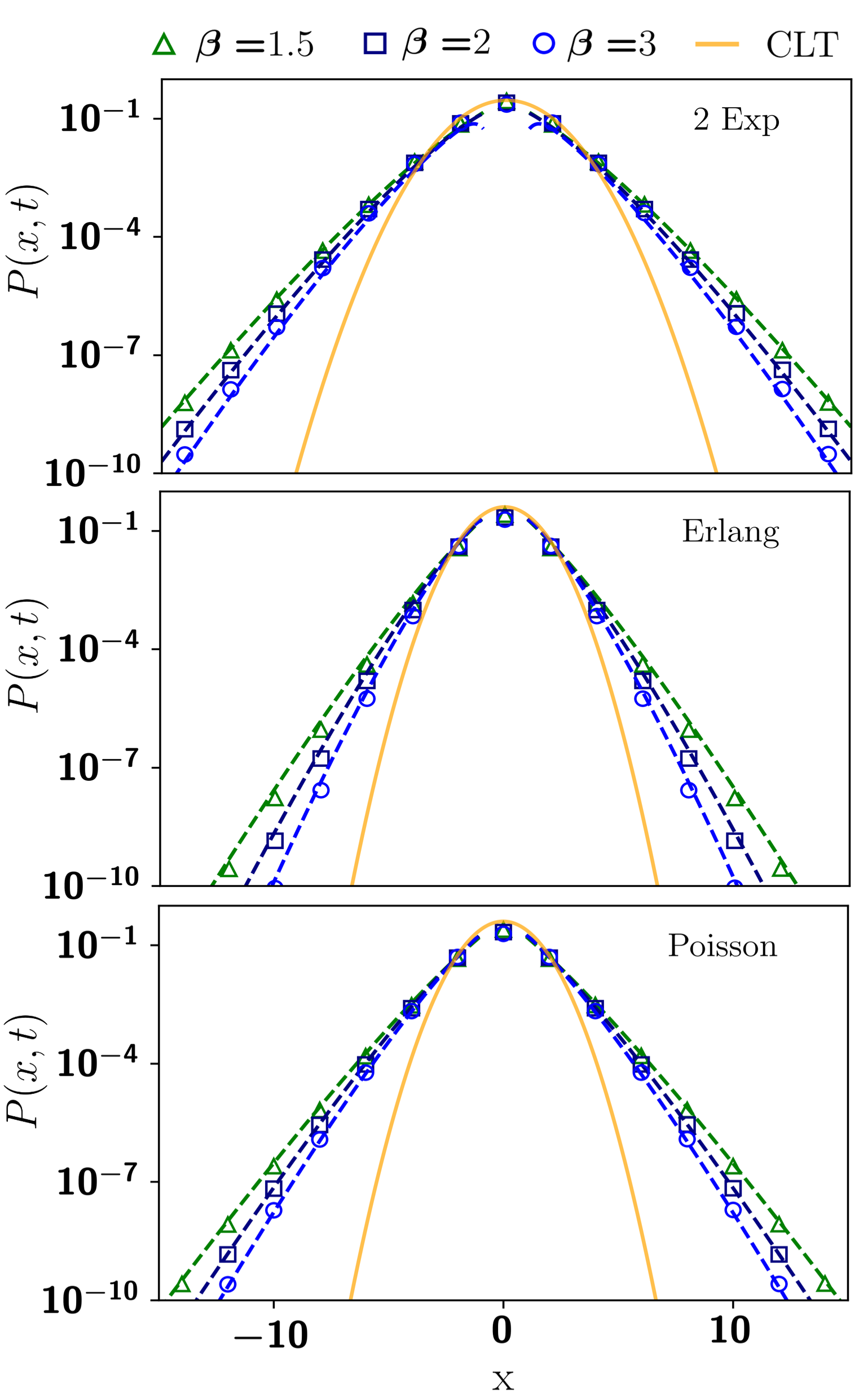}
\caption{
Universality of exponential-like tails. Comparison of $P(x,t)$
obtained from CTRW numerical convolution (symbols), and theoretical predictions; Eq. (\ref{eq_M2:P(x,t)_LDT_Final}) in dashed lines, and CLT in orange lines. The waiting time PDFs are: Eq.~\eqref{eq:sec4Bpsiexmpl} (double exponential) with parameters $\lambda_1 = 1,\lambda_2=3,d_1 =5$ (top), Eq.~\eqref{eq:earlangpsi01} (Erlang) with parameters $k=2,\lambda=4$ (middle), and Eq.~\eqref{eq:psiExponentialClassical} (the classic Poisson case) with $\lambda=1$ (bottom). 
For all those cases, we choose $t/\langle\tau\rangle=1$, i.e., we focus on the scale where a particle jumps on average only once.
Notice the large deviations from the traditional CLT and the nearly exponential tails accurately describing the system without fitting.
Regardless of $\beta$, the tails are approximately exponential.
}\label{Fig_CTRW_Compressed_Exp}
\end{figure}

\subsection{Compressed exponential Jumps in CTRW} \label{sectionwithnumericalcode}
The canonical example for CTRW is Gaussian jumps and exponential waiting times. This example describes various physical systems with well-known solutions in the literature \cite{hamdi2024laplace,barkai2020packets}.
Here, we expand our understanding to a broader range of processes, while still capturing this canonical example, and demonstrating both the Gaussian CLT and the exponential-like tails.
We choose a compressed exponential jump PDF 
\begin{equation}\label{eq01}
f\left( x_i \right) =  \widetilde{N} \exp\left( - \alpha^\beta |x_i|^\beta\right),\quad \beta>1,
\end{equation}
with $\alpha = \sqrt{ \Gamma(3/\beta)/\Gamma(1/\beta)}$, so that $\operatorname{Var}[x_i]=1$ for all $\beta$.
The normalization constant is $\widetilde{N} = \beta \sqrt{\Gamma(3/\beta)}/2 \Gamma^{3/2} (1/\beta)$.
In the large $u$ limit, $\hat{f}(u)$ is approximated using Laplace's method \cite{hamdi2024laplace}:
\begin{eqnarray}\label{eq_M2:hat{f}(u)}
\hat{f}(u) &\approx& \widetilde{N} B_2 |u|^{ \frac{1 }{ 2} \frac{ 2 - \beta }{ \beta -1} }
\exp\left( B_1 |u|^{ \frac{\beta }{\beta-1}} \right),
\end{eqnarray}
See Appendix \ref{ap:CTRW_compressed_exp_calcs} for the explicit form of the constants $B_1, B_2$.
When $\beta=2$, we obtain a Gaussian, and Eq. (\ref{eq_M2:hat{f}(u)}) becomes exact. Using Eq.~\eqref{eq_M2:KofU_CTRW} we derive the saddle point $u_0$ (see Appendix \ref{ap:CTRW_compressed_exp_calcs})
\begin{equation}\label{eq_M2:saddle_point_comressed_exponetial}
u_0 \sim B_3 W_0 \left[  B_4 q ^{ B_5}
\right]^\frac{\beta-1 }{ \beta},
\end{equation}
and the form of the constants $B_3,B_4,B_5$, with $W_0(x)$ being the principal branch of the Lambert function \cite{corless1996lambert}, defined by $W(x) \exp\left[W(x)\right]=x$.
When $\beta>4+2A$, one must use a different branch of the Lambert function and obtain a slightly different solution. A case to be studied in future work.

For large arguments, the Lambert $W$ function can be approximated as $W_0(z) \underset{z\to\infty}{\sim} \ln z - \ln(\ln z)$ \cite{corless1996lambert}. 
However, it should be noted that $\ln z$ is comparable in magnitude to $\ln(\ln z)$ for any practically accessible value of $z$, and we therefore do not advocate using the leading order expression except for illustrative purposes.
Consequently, $u_0$ increases with $x$, but  very slowly.

 Using our saddle point approximation, and Eq. (\ref{eq_M2:P(x,t)_LDT_K(u_0)}) we derive the propagator:
\begin{equation}\label{eq_M2:P(x,t)_LDT_Final}
P(x,t) \approx \frac{\exp \left\{-t\left(\frac{|x|}{t}{\large\chi}_{\scalebox{0.55}{$\mathrm{ARF}$}}\left(\frac{|x|}{t} \right)-\frac{C_{A+1}}{C_A}\right) \right\}}{\sqrt{2\pi |xT^{''}(u_0)|}},
\end{equation}
where
\begin{equation}\label{eq_M2:z(q)}
        {\large\chi}_{\scalebox{0.55}{$\mathrm{ARF}$}}(q)=\frac{B_3 W_0\left[B_4 q^{B_5}\right]-\alpha \left( \frac{2 \beta (\beta - 1)}{4+2A - \beta} \right)^{\frac{1}{\beta}}}{W_0\left[B_4 q^{B_5}\right]^{\frac{1}{\beta}}},
\end{equation}
and the constants $B_3,B_4,B_5$ are provided by EQ.~\eqref{eq:Bconstantsappendix}.
Since ${\large\chi}_{\scalebox{0.55}{$\mathrm{ARF}$}}\left(q\right)$ is slowly varying, being a ratio of Lambert $W$ functions that for large arguments grow only logarithmically, we recover Laplace’s first law, which predicts exponential-like decay in the tails of the PDF.

When $q\to\infty$, Eqs. (\ref{eq_M2:P(x,t)_LDT_Final},\ref{eq_M2:z(q)}) yield
\begin{eqnarray}\label{eq_M2:Z(q)_and_P(x,t)_asymp}
    {\large\chi}_{\scalebox{0.55}{$\mathrm{ARF}$}}(q)&\sim& B_3\ln(B_5q)^{1-1/\beta}\nonumber\\
    P(x,t)&\simeq& \exp{\left[-|x|B_3\ln\left(B_5\frac{|x|}{t}\right)^{1-1/\beta}\right]}.
\end{eqnarray}
Namely, our predicted exponential-like Laplace tails \cite{barkai2020packets}, with logarithmic corrections.
While this prediction is aesthetically pleasing, we prefer to use the full Lambert solution in Eq. (\ref{eq_M2:z(q)}), which is significantly more accurate for any measurable value of $x/t$.
We also note that Eq. (9) in \cite{barkai2020packets} contains an error; the correct expression is Eq. (\ref{eq_M2:z(q)}) developed above~\cite{correctionbarkaiburov}.

In Fig. \ref{Fig_CTRW_Compressed_Exp}, we plot our results for three different values of $\beta; 1.5,2,3$, and three different waiting times PDFs; Eq.~\eqref{eq:sec4Bpsiexmpl} (double exponential) with parameters $\lambda_1 = 1,\lambda_2=3,d_1 =5$, Eq.~\eqref{eq:earlangpsi01} (Erlang) with parameters $k=2,\lambda=4$, and Eq.~\eqref{eq:psiExponentialClassical} (the classic Poisson case) with $\lambda=1$. 
For all those cases, we choose $t/\langle\tau\rangle=1$, i.e., we focus on a time scale where a particle jumps on average only once. Our theory matches all those systems, with pronounced and measurable Laplace (exponential-like) tails.

\begin{figure}[t]
\centering
\includegraphics[width=0.485\textwidth]{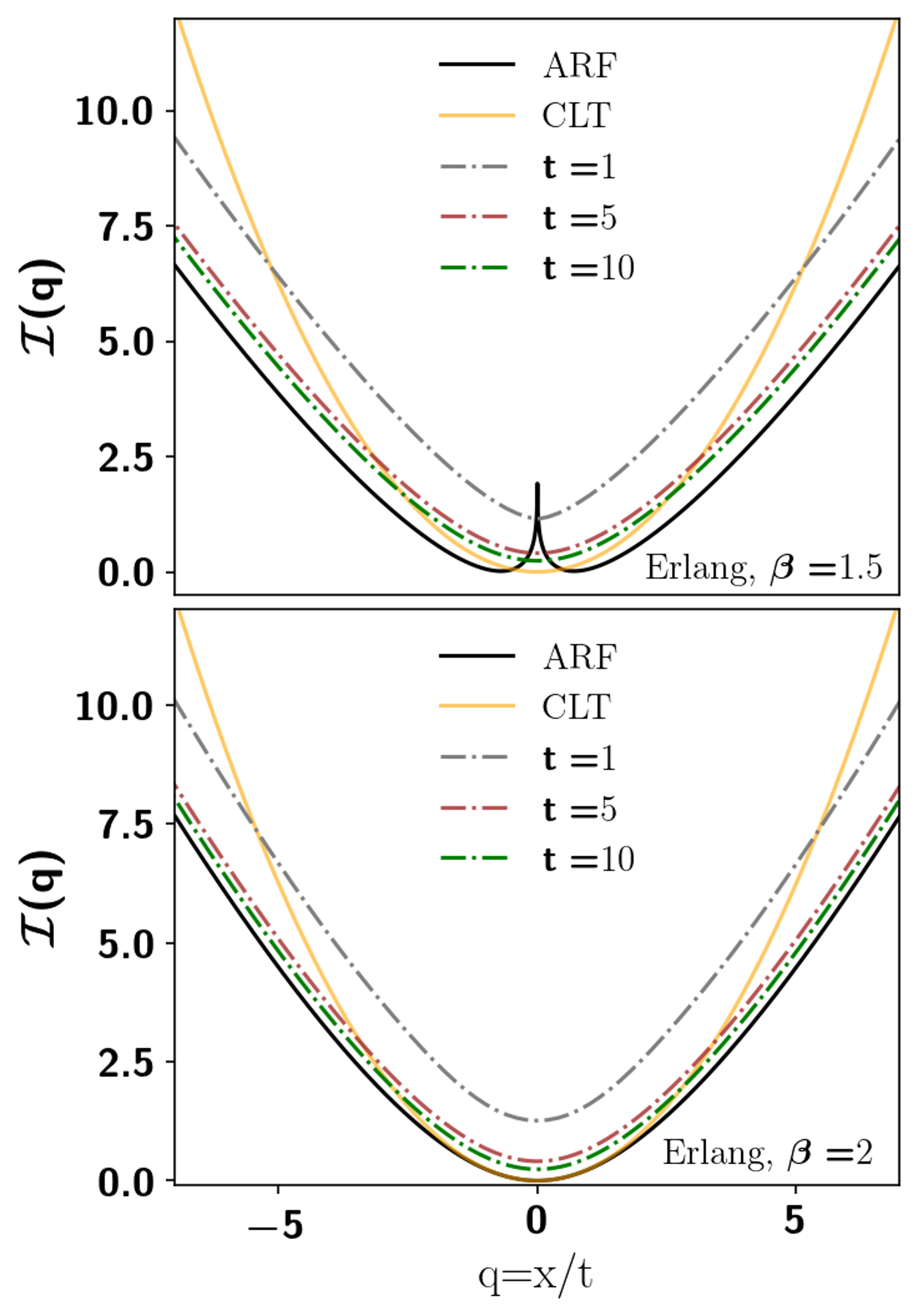}
\caption{
Using numerical solution of the CTRW presented in Sec.~\ref{sectionwithnumericalcode}, we plot the rate function for $P(x,t)$. For three finite $t$ we plot $-\log[P(x,t)]/t$ versus $q=x/t$. 
The numerical results are presented by dashed lines while analytical form of the rate function is provided for the CLT approximation via Eq.~\eqref{eq:pxtCLTFirst} (orange line) and for the ARF via Eq.~\eqref{eq_M2:P(x,t)_LDT_Final} (black line). For the PDF of the waiting times we used Eq.~\eqref{eq:earlangpsi01} (Erlang) with parameters $k=2,\lambda=4$, while the jump distribution follows Eq.~\eqref{eq01} with $\beta=1.5,2$.
For large enough $t$, the ARF is the dominating behavior in the tails, while the CLT is dominating in the center.
}\label{Fig_CTRW_RF_Erlang}
\end{figure}

Finally, we use the rate function formalism to compare our results to numerical simulation of the CTRW process. 
The rate function is usually defined as 
 the limit of $-\log\left[P(X,t)\right]/t$ while $t\to\infty$. 
 In the case when $P(x,t)$ follows the form $P(x,T)\sim\exp\left[-t \mathcal{I}(x/t)\right]$, the function $\mathcal{I}\left(q=\frac{x}{t}\right)$ is the rate function. 
 From  Eq.~\eqref{eq:pxtCLTFirst} the CLT approximation for $\mathcal{I}(q)$ can be obtained, while the ARF approximation is obtained from Eq.~\eqref{eq_M2:P(x,t)_LDT_Final}. 
In Fig. \ref{Fig_CTRW_RF_Erlang}, we present the behavior of $\mathcal{I}(q)$ as obtained for a series of growing $t$. 
As $t$ is growing a convergence to the limit form is observed. 
It is important to notice that the behavior of the tails, as observed from the rate function, are exponential-like for all finite $t$ that we explored.  

The numerical algorithm for obtaining the representation of $P(x,t)$ is as follows: 
In the discussed examples we use waiting times for which $Q_t(n)$ is known exactly, leaving us with the tasks of numerically evaluating $\phi(x\mid n)$ and summing the truncated series in Eq. (\ref{eq_M2:P(x,t)_summation}). This approach avoids the difficulties associated with the numerical Fourier or Laplace transforms.
Since $\phi(x\mid n)$ is a sum of $n$ IID random variables, we compute it via numerical convolution, controlling the numerical error by adjusting the resolution of the discretization used in the convolution procedure. The basic code implementing this calculation is provided in Appendix \ref{ap:Numerical_convolution}.

\section{Rate function formalism}
In a recent work of Pacheco-Pozo and Sokolov \cite{PhysRevE.103.042116}, the authors presented a general expression for the large deviation rate function in CTRWs in terms of the corresponding rates of the distributions of steps’ lengths and waiting times.
In their technique, one takes the two rate functions $\mathcal{\widetilde{I}}(\tilde{y}=t/n)$, $\mathcal{R}(\tilde{q}=x/n)$ of $Q_t(n)\asymp \exp\left[-n \mathcal{\widetilde{I}}(\Tilde{y})\right]$,  $\phi(x|n)\asymp\exp\left[-n \mathcal{R}(\tilde{q})\right]$, and construct the full rate function of the propagator $P(x,t) \asymp \exp\left[-t \mathcal{\chi}(q=x/t)\right]$, using the relation \cite{PhysRevE.103.042116}:
\begin{eqnarray}\label{eq_M2:rateFuncFormalizm}
    \mathcal{\chi}(q=x/t)=-\underset{\zeta}{\rm{sup}}\left[-\frac{1}{\zeta}\left(\mathcal{R}(q\zeta)+\mathcal{\widetilde{I}}(\zeta)\right)\right].
\end{eqnarray}
While this relation is very general, it is not always easy to analytically solve, since $\mathcal{\widetilde{I}}(\Tilde{y})$, and $\mathcal{R}(\tilde{q})$ do not necessarily admit closed form expressions.
Moreover, it assumes large $t$, and neglects all the pre-factors of $P(x,t)$, hence missing some information of the propagator.

In this section, we aim to implement their methods on our asymptotic results for $Q_t(n)$ and $\phi(x|n)$, and check if one can apply this technique to the asymptotic rate functions (ARFs) and still obtain meaningful results in the corresponding limit.
Namely, we compare this combined method to the one used in the text, Eq. (\ref{eq_M2:P(x,t)_LDT_Final}), for describing the Laplace tails in the large $x$ and large $t$ limit.
We do so while using our ARF approximation from Eq. (\ref{eq_M2:QtN_RF_APPROX_GENERAL}), for the temporal rate, where, in our language, $\mathcal{\widetilde{I}}(\tilde{y}=t/n) =\tilde{y}\mathcal{I}(1/\tilde{y}=n/t) $.
\subsection{Gaussian jumps}
We apply this theory to the case of Gaussian jumps, using our waiting-time asymptotic rate function approximation in Eq. (\ref{eq_M2:QtN_RF_APPROX_GENERAL}). Here, we have $\beta=2$, and the spatial rate function is
\begin{eqnarray}\label{eq_M2:RF_approx_x/n_gauusian}
        \mathcal{R}(\tilde{q}=x/n) = \frac{\tilde{q}^2}{2},
\end{eqnarray}
which is exact for every $\tilde{q}$. Inserting the rates from Eqs. (\ref{eq_M2:QtN_RF_APPROX_GENERAL},\ref{eq_M2:RF_approx_x/n_gauusian}) into Eq. (\ref{eq_M2:rateFuncFormalizm}), we get the large $n$, i.e., large $q=x/t$ approximation for the rate function $\chi(q)$
\begin{eqnarray}\label{eq_M2:RF_chi_Gaussian}
    &&\mathcal{\chi}(q)\approx-\underset{\zeta}{\rm{sup}}\nonumber\\
    &&\left[-\frac{1}{\zeta}\left\{
    \frac{(\zeta  z)^{2}}{2} - (A+1) \log \left(\frac{e \left(A! C_A\right){}^{\frac{1}{A+1}}}{A+1}\zeta \right)-\frac{  C_{A+1}}{C_A}\zeta
    \right\} \right],\nonumber\\
\end{eqnarray}
with the minima located at
\begin{eqnarray}
    \zeta^* = \frac{\sqrt{(A+1) W_0\left(\chi_0  q^2\right)}}{z}\ ,\ \ \chi_0 =\frac{A+1}{\left(A! C_A\right){}^{\frac{2}{A+1}}}.
\end{eqnarray}
Inserting back to Eq. (\ref{eq_M2:RF_chi_Gaussian})
\begin{eqnarray}\label{eq_M2:RF_chi_beta_is_2}
     \chi(q)\approx\sqrt{A+1}  \left(\sqrt{W_0\left(\chi_0 q^2\right)}-\frac{1}{\sqrt{W_0\left(\chi_0 q^2\right)}}\right)q-\frac{C_{A+1}}{C_A}.\nonumber\\
\end{eqnarray}
This result perfectly aligns with Eq. (\ref{eq_M2:z(q)}) when $\beta=2$. We note that using this method does not yield the pre-factor in Eq. (\ref{eq_M2:P(x,t)_LDT_Final}); thus, it cannot be used as a good approximation for the propagator $P(x,t)$ for finite $t$.
\subsection{General jumps}
Now we aim to derive the universal Laplace tails, which arise for $\beta>1$ (see Eq.~\eqref{eq01}), using the rate function formalism for the approach developed in this work (ARF) and in~\cite{PhysRevE.103.042116} via Eq.~\eqref{eq_M2:rateFuncFormalizm}.
This is achieved by first obtaining an asymptotic approximation of $\phi(x|n)$ using the LDT formalism, with the same choice of jump distribution as in Eq. (\ref{eq01}) for $\beta>1$.
Taking the inverse Fourier transform of Eq. (\ref{eq_M2:tilde{f}^n}), while changing variable $i k =u$, reads
\begin{eqnarray}
\phi(x|n) &=& \frac{1}{2 \pi i} \int_{- i \infty} ^{i \infty} \exp{\left[ - x\left(u-\frac{n}{x}\ln\left[\hat{f}( u)\right]\right) \right]}  {\rm d} u.\nonumber\\
\end{eqnarray}
Using the saddle point approximation in the large $x$ limit
\begin{eqnarray}\label{eq_M2:phi_x_n_general}
    \phi(x|n)&\approx& \frac{\exp\left[-x\mathcal{R}(\tilde{q},u_0)\right]}{\sqrt{2 \pi \left|n \frac{\partial^2 \ln[\hat{f}(u)]}{\partial u^2}\big|_{u_0}\right|}},
\end{eqnarray}
where we denoted
\begin{eqnarray}
    \mathcal{R}(\tilde{q},u_0) &=& u_0-\frac{1}{\tilde{q}}\ln\left[\hat{f}(u_0)\right].\label{eq_M2:phi_x_n_rate_function_general}
\end{eqnarray}
Here, $u_0$ is the saddle point, and $\tilde{q}=x/n$.
To find $u_0$, we take the derivative of $\mathcal{R}$ and set to zero.
In the large $x$ limit, $u_0$ increases with $\tilde{q}=x/n$, and we approximate
\begin{eqnarray}
    \ln\left[\hat{f}(u)\right] &\approx&
    B_1 u^{ \frac{\beta }{\beta-1}}+\ln\left[\widetilde{N} B_2 u^{ \frac{1 }{ 2} \frac{ 2 - \beta }{ \beta -1} }\right].\label{eq_M2:log_of_f_of_u_approx}
\end{eqnarray}
We neglect the logarithmic correction in Eq. (\ref{eq_M2:log_of_f_of_u_approx}) for the calculation of $u_0$, and obtain
\begin{eqnarray}
    u_0 \approx \left(\frac{(\beta-1)   }{\beta B_1}\tilde{q}\right)^{\beta-1}.\ \
\end{eqnarray}
Hence,
\begin{eqnarray}
    \ln\left[\hat{f}(u_0)\right] &\approx&
    B_1 \left(\frac{(\beta-1) }{\beta B_1}\tilde{q}\right)^\beta+
    \ln \left[\widetilde{N}B_2  \left(\frac{(\beta-1) }{\beta B_1}\tilde{q}\right)^{1-\frac{\beta}{2}}\right].\nonumber\\
\end{eqnarray}
Substituting into $\mathcal{R}(\tilde{q},u_0)$, yields;
\begin{eqnarray}\label{eq_M2:RF_approx_x/n}
    \mathcal{R}(\tilde{q},u_0) &=&
    \left(\alpha \tilde{q}\right)^{\beta} + \frac{1}{2}(\beta - 2)\ln\left(\tilde{q}\right) + R_0,
    \nonumber\\
    R_0&=&\ln\left(\frac{1}{\widetilde{N}} \sqrt{\frac{(\beta - 1)\beta}{2\pi}\alpha^{\beta}}\right).
\end{eqnarray}
When $\beta=2$, we recover the Gaussian form, and all the provided results in this section become exact.
The leading result matches Eq. (1) in \cite{barkai2020packets}. The added corrections vanish when $\beta=2$ (since $R_{0}|_{\beta=2}=0$), and we recover Eq. (\ref{eq_M2:RF_approx_x/n_gauusian}).
Inserting the rates from Eqs. (\ref{eq_M2:QtN_RF_APPROX_GENERAL},\ref{eq_M2:RF_approx_x/n}) into Eq. (\ref{eq_M2:rateFuncFormalizm}), we get the large $x$, and $n$ approximation, namely the rate function that describes the tails of the distribution
\begin{eqnarray}\label{eq_M2:RF_chi}
    \mathcal{\chi}\left(q=\frac{x}{t}\right)\approx&-&\underset{\zeta}{\rm{sup}}
    \Bigg[\frac{1}{\zeta}\Bigg\{
    (\alpha  \zeta  q)^{\beta }+\frac{1}{2} (\beta -2) \log (\zeta  q)\nonumber\\
    &-&(A+1) \log (\zeta )+r_0-\frac{  C_{A+1}}{C_A}\zeta
    \Bigg],\nonumber\\
    r_0 &\equiv& R_0+(A+1) \log \left(\frac{(A+1) }{e\left(A! C_A\right)^{\frac{1}{A+1}}}\right)
\end{eqnarray}
Taking a derivative with respect to $\zeta$, and comparing to zero to find the minima, we get the following equation:
\begin{eqnarray}
&&2 (A+1) \log (\zeta )+2 (\beta -1) (\alpha  \zeta  q)^{\beta }\nonumber\\
&-&(\beta -2) \log (\zeta  q)-4-2 A+\beta -2 r_0 =0,
\end{eqnarray}
which has an exact solution using the Lambert function:
\begin{eqnarray}\label{eq_M2:sokolov_RF_zeta_star}
    \zeta^*(q) 
    &=&\frac{1}{\zeta_0^{1/\beta } q}W_0\left(\zeta_0 \zeta_1 q^{B_5}\right)^{1/\beta },
    \nonumber\\
    \zeta_0 &\equiv& \frac{2 (\beta -1) \beta  \alpha ^{\beta }}{2 A-\beta +4}\ ,\quad \zeta_1 \equiv e^{\frac{2 \beta  r_0}{2 A-\beta +4}+\beta }.
\end{eqnarray}
Inserting back to the rate function in Eq. (\ref{eq_M2:RF_chi}), and simplifying leads to
\begin{eqnarray}\label{eq_M2:sokolov_RF_general_beta}
    \chi(q)\approx&-&\frac{C_{A+1}}{C_A}-\frac{1}{\zeta^*(q)}
    \Bigg\{(\alpha  z\zeta^*(q) )^{\beta }\nonumber\\
    &+& \frac{\beta-2 A -4}{2} \log (\zeta^*(q))+\frac{\beta -2}{2} \log (q)+r_0
    \Bigg\}.\nonumber\\
\end{eqnarray}
For $\beta = 2$, the above equation reduces to Eq. (\ref{eq_M2:RF_chi_beta_is_2}). While closely related, this result is not identical to that in Eq. (\ref{eq_M2:z(q)}), owing to the different approximation used for $\phi(x|n)$. Nevertheless, Fig. \ref{Fig_RF_compare} shows that the two results exhibit almost perfect agreement, and in Appendix \ref{ap:Convergence_sokolov_to_us} we present the convergence of the two solutions for large $q$. Furthermore, when one approximates the Lambert function as a logarithm, one recovers the same scaling behavior of the rate function (see Appendix \ref{ap:leading_scaling_sokolov}), namely the $\sim q \log(q)^{1-1/\beta}$ scaling.
We note that, since the logarithm is a poor approximation of the Lambert $W$ function, one cannot compare the coefficients of the leading behavior using this method; only the leading scaling of the function can be compared, which matches in the limit $q \to \infty$.
\begin{figure}[ht]
\centering
\includegraphics[width=0.45\textwidth]{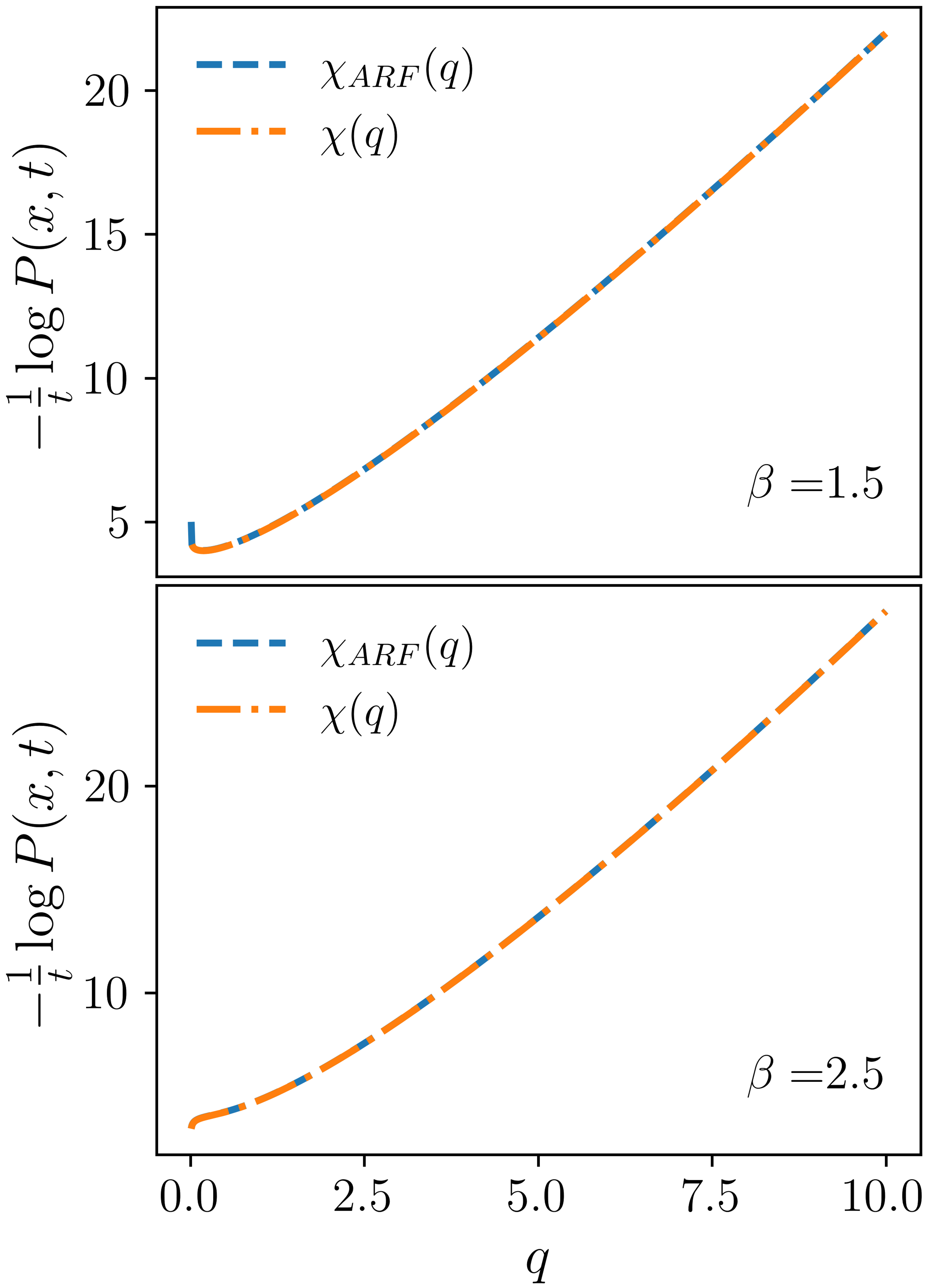}
\caption{
Comparison of the rate function of $P(x,t)$ obtained by our method using ${\large\chi}_{\scalebox{0.55}{$\mathrm{ARF}$}}(q)$ from Eq. (\ref{eq_M2:z(q)}) (blue dashed) and by the formalism of Ref. \cite{PhysRevE.103.042116}, described by Eq. (\ref{eq_M2:sokolov_RF_general_beta}) (orange dash-dotted). The parameters are: $A=5$, $C_A=1$, $C_{A+1}=-5$, with $\beta=1.5$ (top) and $\beta=2.5$ (bottom). For both values of $\beta$, we see a good fit between the two approaches.
}\label{Fig_RF_compare}
\end{figure}

\section{Discussion}
This work examines the use of large deviations theory and saddle point approximations in describing diffusive motion arising from continuous-time random walks whose waiting-time distributions are analytic in the short time limit. LDT is typically linked to the study of rare events and its rate functions are often regarded as challenging to sample numerically \cite{Lecomte_2007,PhysRevLett.96.120603,debiossac2023convergence,Thapa_2021}.
We study two limits. The large $x$ and finite $t$, and the LDT limit of large $x$ and large $t$ keeping the ration $x/t$ constant. 
Our results are suitable for finite $t$ comparison to experimentally and numerically obtained data.
To substantiate this claim, we systematically compare our analytical approximations with direct numerical computations, employing the numerical methods described in Section \ref{sectionwithnumericalcode}. 
By examining finite observation times and large values of $x$, we show that the nearly exponential decay of both $P(x,t)$ and $Q_t(n)$, together with the accompanying logarithmic corrections, is more accessible to empirical sampling than previously thought. 
This is especially relevant for single-tracer experiments, where the microscopic time scale set by the mean time between jumps can be comparable to the total observation time $t$, allowing the exponential-like tails of $P(x,t)$ and $Q_t(n)$ to be directly observed.

Focusing on $Q_t(n)$, the probability that exactly $n$ renewals occur during time $t$, we elaborated on the three main regimes exhibited by this observable: the “rare event” regime at \textit{large} $y=n/t$, the central limit theorem regime, and the “rare event” regime at \textit{small} $y=n/t$. The first yields universal Laplace (exponential-like) tails, the second produces the familiar Gaussian form, and the third reflects the slowest-decaying component of the waiting-time distribution, which dominates the behavior in this limit.


For $P(x,t)$ of CTRW we focus on two regimes of the corresponding rate function $\mathcal{I}\left(q=x/t\right)$.
The well-known central limit regime for \textit{small} $x/t$ where $\mathcal{I}\left(q\right)\sim q^2$ and the ``rare-events'' (Laplace) regime for \textit{large} $q$ where $\mathcal{I}\left(q\right)$ is linear with logarithmic corrections. 
Here, we extend the understanding developed in earlier works \cite{chaudhuri2007universal,barkai2020packets,PhysRevE.103.042116} by providing a complete description of $P(x,t)$ in the form of a closed analytical expression, valid for a very general class of waiting time and jump distributions. 
In doing so, Eq.~\eqref{eq_M2:z(q)} also corrects an error that appeared in a constant in Eq.~(9) in \cite{barkai2020packets}. 
The underlying physical picture of exponential decay of $P(x,t)$, presented in~\cite{barkai2020packets}, remains unchanged. 
The present formulation shows that the regime of Laplace tails is a robust feature of CTRW transport, describing rare fluctuations in the same way that the Gaussian regime describes typical fluctuations near the center.

{\bf Acknowledgements:}
    We acknowledge the support of the Israel Science Foundation's grants 2311/25, and 3791/25.
\onecolumngrid
\appendix
\counterwithin{figure}{section}%
\section{From Eq. (\ref{eq_M2:P(k,t)_exact}) to Eq. (\ref{eq_M2:P(k,t)_large_n_limit})}\label{ap:from_exact_p(k,t)_to_large)_n_limit}
Here, we provide a mathematical justification for the transition from Eq. (\ref{eq_M2:P(k,t)_exact}) to Eq. (\ref{eq_M2:P(k,t)_large_n_limit}).
Starting with the leading term of $Q_t(n)$ in Eq. (\ref{eq_M2:Q_T_n_large_y_LDT_Solution}): 
\begin{eqnarray}
        Q_t(n)= \frac{\left(c_A A!t^{A+1}\right)^n}{(n(A+1))!}\exp\left[\frac{C_{A+1}}{C_A}t\right].
\end{eqnarray}
we insert to Eq. (\ref{eq_M2:P(k,t)_exact})
\begin{eqnarray}\label{eq_M2:P(k,t)_Q_t(n)}
\widetilde{P} (k, t) &=& \sum_{n=0} ^\infty \frac{\left\{\left[C_A A!\right]^{1/(A+1)} t\right\}^{n(A+1)}}{(n(A + 1))!} {\rm{e}}^{\frac{C_{A+1}}{C_A}t}\widetilde{f}^n(k)+\sum_{n=0} ^\infty \delta(n)\widetilde{f}^n(k)\nonumber\\
&=&{\rm{e}}^{\frac{C_{A+1}}{C_A}t}\sum_{n=0} ^\infty \frac{\left\{\left[C_AA!\widetilde{f}(k)\right]^{1/(A+1)} t \right\}^{n(A+1)}}{(n(A + 1))!}+\sum_{n=0} ^\infty \delta(n)\widetilde{f}^n(k).
\end{eqnarray}
Here, $\delta(n)$ is the error part due to the approximation of $Q_t(n)$.
Since $\delta(n)$ is a strongly decaying function, we can consider the sum $\sum_{n=0} ^\infty \delta(n)\widetilde{f}^n(k)$ as a finite number of correction terms, hence when inverse transforming back to $x$ space, and taking $x\to \infty$, we can neglect this contribution. This means that the approximation made here is for small values of $k$.
Changing variables for the running index $n = n(A+1)$, and considering only the first term in the correction series:
\begin{eqnarray}
\widetilde{P} (k, t) &=& {\rm{e}}^{\frac{C_{A+1}}{C_A}t}\sum_{n=0} ^\infty \frac{\left\{\left[C_A A!\widetilde{f}(k)\right]^{1/(A+1)} t \right\}^n}{n!}\nonumber\\
&=& \exp\left[ \frac{C_{A+1}}{C_A}t+\left[C_A A!\widetilde{f}(k)\right]^{1/(A+1)} t\right].
\end{eqnarray}
Which is Eq. (\ref{eq_M2:P(k,t)_large_n_limit}) of the main text.
\section{Exact Solution For the sum of two exponentials}\label{ap:2Exp_Exact_sol}
In this appendix, we outline how to get an exact solution for the PDF $Q_t(n)$, starting with the waiting time PDF $\psi(\tau) = d_0\left(d_1 {\rm{e}}^{-\lambda_1 \tau}+ {\rm{e}}^{-\lambda_2 \tau}\right)$, and it's Laplace transform $\hat{\psi}(s) = \frac{d_0 d_1}{s+\lambda_1} + \frac{d_0}{s+\lambda_2}$. The main idea is to use the residue method and insert into Mathematica to get a closed analytical form. 
We start by writing $\hat{\psi}(s)$ as 
\begin{eqnarray}
    \hat{\psi}(s)= \frac{g_1(s)}{(s+\lambda_1)(s+\lambda_2)},
\end{eqnarray}
with $g_1(s) = d_0d_1(s+\lambda_2)+d_0(s+\lambda_1)$, and
\begin{eqnarray}
    \frac{1-\hat{\psi}(s)}{s}= \frac{g_2(s)}{s(s+\lambda_1)(s+\lambda_2)},
\end{eqnarray}
with $g_2(s) = s^2 + s \left( \lambda_1 + \lambda_2 - d_0d_1 - d_0 \right) + \lambda_1 \lambda_2 - \lambda_1 d_0 - \lambda_2 d_0d_1$.
Inserting to Eq. (\ref{eq_M2:laplacetrans_Q_t(n)}), and taking the inverse Laplace transform gives
\begin{eqnarray}
    Q_t(n) &=&\frac{1}{2\pi i}\int \underbrace{e^{st}\frac{g_1(s)g_2(s)^n}{s(s+\lambda_1)^{n+1}(s+\lambda_2)^{n+1}}}_{h(s)}ds,
\end{eqnarray}
which can be solved exactly using the residue method around our three distinct singularities at $s^*\in\{0,-\lambda_1,-\lambda_2\}$, i.e.,
\begin{eqnarray}\label{eq_M2:Q_t_n_SumOfTwoExp}
    Q_t(n) &=& {\rm Res}\{h(s),s=0\}+{\rm Res}\{h(s),s=-\lambda_1\}+{\rm Res}\{h(s),s=-\lambda_2\},
\end{eqnarray}
which can be calculated exactly by inserting the following expressions into Wolfram Mathematica:
\begin{eqnarray}
    \text{res}_1 &=& \left.\frac{g_1(s)\, g_2(s)^n}{(s+\lambda_1)^{n+1}(s+\lambda_2)^{n+1}}\right|_{s=0} \nonumber\\[6pt]
    \text{res}_2 &=& \left.\frac{1}{n!} \frac{d^n}{ds^n} \left[ \frac{1}{s(s+\lambda_2)^{n+1}}\, e^{st} \, g_1(s)\, g_2(s)^n \right] \right|_{s=-\lambda_1} \nonumber\\[6pt]
    \text{res}_3 &=& \left.\frac{1}{n!} \frac{d^n}{ds^n} \left[ \frac{1}{s(s+\lambda_1)^{n+1}}\, e^{st} \, g_1(s)\, g_2(s)^n \right] \right|_{s=-\lambda_2}.
\end{eqnarray}
\section{Deriving $u_0$}\label{ap:CTRW_compressed_exp_calcs}
Starting with the definition of $T(u)$ in Eq. (\ref{eq_M2:KofU_CTRW}), we take the first derivative and compare to zero to get our equation for $u_0$:
\begin{eqnarray}
&&\frac{1}{2} (\frac{1}{(A+1) q}\left(\frac{\pi }{2}\right)^{\frac{1}{2 A+2}} (\beta-1)^{-\frac{1}{2 (A+1)}-1} \beta^{\frac{2 \beta-3}{2 (A+1) (\beta-1)}} (\text{CA} A!)^{\frac{1}{A+1}} u^{\frac{-2 A (\beta-1)-3 \beta+4}{2 (A+1) (\beta-1)}} \Gamma \left(\frac{1}{\beta}\right)^{\frac{(2 A-3) \beta+6}{4 (A+1) (\beta-1)}} \nonumber\\
&&\Gamma \left(\frac{3}{\beta}\right)^{\frac{\beta-2}{4 (A+1) (\beta-1)}} \left((\beta-2) \Gamma \left(\frac{1}{\beta}\right)^{\frac{\beta}{2-2 \beta}}-2 (\beta-1) \beta^{\frac{1}{1-\beta}} u^{\frac{\beta}{\beta-1}} \Gamma \left(\frac{3}{\beta}\right)^{\frac{\beta}{2-2 \beta}}\right)\nonumber\\
&&\exp \left(\frac{(\beta-1) \beta^{\frac{\beta}{1-\beta}} u^{\frac{\beta}{\beta-1}} \left(\frac{\Gamma \left(\frac{1}{\beta}\right)}{\Gamma \left(\frac{3}{\beta}\right)}\right)^{\frac{\beta}{2 (\beta-1)}}}{A+1}\right)+2 )=0.
\end{eqnarray}
Here, there are two solutions, and we keep only the one corresponding to the large-$x$ limit, i.e., neglect the $(\beta-2) \Gamma \left(\frac{1}{\beta}\right)^{\frac{\beta}{2-2 \beta}}$ term in the parenthesis. Identifying that the equation is of the form $1-A_1 \exp\left[A_2u^{\beta/\beta-1}\right]u^{A_3}$, we utilize the Lambert function $W(x)$ \cite{corless1996lambert} which satisfies $W(x) \exp\left[W(x)\right]=x$. The branch of the Lambert function relevant to our study is the well documented principal branch $W_0 (x)$.
\begin{eqnarray}
u_0 \sim B_3 W_0 \left[  B_4 q ^{ B_5}\right]^\frac{\beta-1 }{ \beta} .
\end{eqnarray}
For large values of the argument, the Lambert function is expressed in terms of logarithmic functions, i.e., $W_0 (z) \underset{z\to\infty}{\sim} \ln(z) - \ln(\ln(z))$,
 hence $u_0$ is growing with $x$, however very slowly. This holds only when $\beta<4+2A$, while the opposite case demands further study. 
The constants are:
\begin{eqnarray}
\label{eq:Bconstantsappendix}
    B_1 &=& (\beta-1)(\beta \alpha)^{\frac{\beta}{1-\beta}},\nonumber\\
    B_2 &=& \sqrt{2\pi/(\beta-1)(\beta\alpha^{\beta})^{\frac{1}{\beta-1}}},\nonumber\\
     B_3&=& \beta \alpha \left( \frac{2\beta(\beta - 1)}{4+2A - \beta} \right)^{\frac{1 - \beta}{\beta}}, \nonumber\\ 
     B_4&=&\frac{(2 (\beta-1) \beta) (\beta \alpha)^{-\frac{\beta}{\beta-1}} \left(\frac{\beta^{\frac{2 A-2 \beta+5}{-2 A \beta+2 A-2 \beta+2}} \left(\frac{\pi }{2 (\beta-1)}\right)^{\frac{1}{2 A+2}} \Gamma \left(\frac{1}{\beta}\right)^{\frac{(2 A-3) \beta+6}{4 (A+1) (\beta-1)}} \Gamma \left(\frac{3}{\beta}\right)^{\frac{2 A \beta+\beta+2}{-4 A \beta+4 A-4 \beta+4}} (C_A A!)^{\frac{1}{A+1}}}{A+1}\right)^{-B_5}}{2 A-\beta+4},\nonumber\\
     B_5
     &=&\frac{ 2(1+A) \beta  }{ 4+2A-\beta}.
\end{eqnarray}
Using $\exp[ a W(x)] = (x)^a / [ W(x)]^a$ to calculate $\hat{f}(u_0)$, we obtain $T(u_0) = u_0 - \frac{1}{q} \left[ \frac{C_{A+1}}{C_A}+\left[C_A A!\hat{f}(u_0)\right]^{1/(A+1)} \right]$. 
\section{Code for Numerical convolution}\label{ap:Numerical_convolution}
Python code for numerically generating $\phi(x|n)$:
\begin{verbatim}
alphy = lambda b: np.sqrt(gamma(3 / b) / gamma(1 / b))
norm = lambda b: 0.5 * b * np.sqrt(gamma(3 / b)) / gamma(1 / b) ** (3 / 2)
pdf_single_jump = lambda x, b: norm(b) * np.exp(-(abs(alphy(b) * x)) ** b)

def convolve_up_to_N(y_arr, xVals_N, N):
    """ Returns an array of the first N convolutions. """
    dx = xVals_N[1] - xVals_N[0]
    convolutions = [np.zeros(len(y_arr), dtype=int), y_arr]
    for i in tqdm(range(1, N - 1)):
        convolutions.append(dx * np.convolve(convolutions[i], y_arr, mode='same'))
    return convolutions

N_steps_ = 100
bd = 200
xVals = np.linspace(-bd, bd, 2 * bd * 10 ** 2)
np.save('FILENAME.npy',
        convolve_up_to_N(pdf_single_jump(xVals,3), xVals, N_steps_))
\end{verbatim}
Python code for summing the series in Eq. (\ref{eq_M2:P(x,t)_summation}):
\begin{verbatim}
def SUM_pdf_x_t(Q_t_n, P_N_x, t):
    sum = np.zeros(len(P_N_x[1]))
    for n in range(1, len(P_N_x)):
        sum = sum + Q_t_n(t, n) * P_N_x[n]
    return sum
\end{verbatim}
\section{Applying the asymptotic solution of $P(x,t)$ for finite $x$}\label{ap:P(x,t)_Laplace_Erlang_mismatch}
In this appendix, we demonstrate why using the asymptotic expression for $P(x,t)$ in Eq. (\ref{eq_M2:Z(q)_and_P(x,t)_asymp}) is not sufficient when one wishes to adequately describe the tail profile of the PDF $P(x,t)$.
In Fig \ref{fig_P(x,t)_Laplace_Erlang_mismatch}, we demonstrate, for the Erlang distribution and $\beta=1.5,3$, a replication of Fig \ref{fig_P(x,t)_Laplace_Erlang_mismatch}, now with Eq. (\ref{eq_M2:Z(q)_and_P(x,t)_asymp}), to demonstrate why we do not advocate for such approximation. It is clear that while the full Lambert solution fits well with the data for finite $|x|$, the logarithmic approximation in Eq. (\ref{eq_M2:Z(q)_and_P(x,t)_asymp}) does not.
\begin{figure}[ht]
\centering
\includegraphics[width=0.5\textwidth]{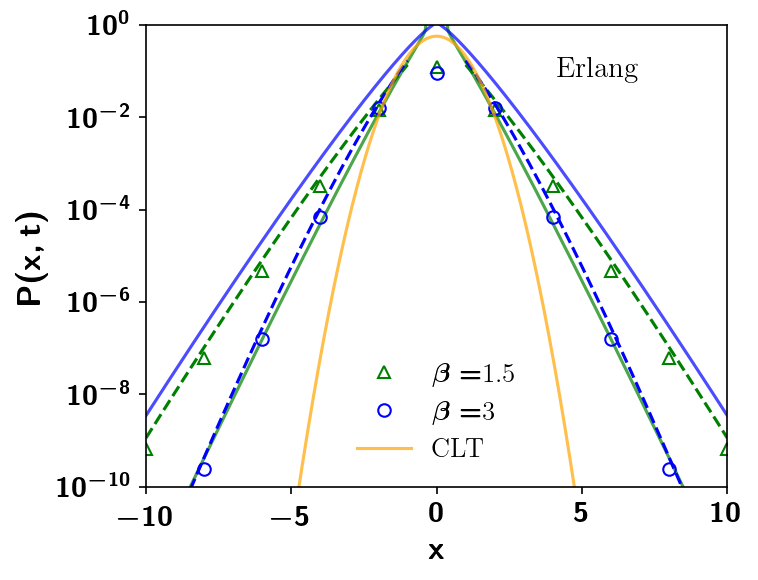}
\caption{Comparison of $P(x,t)$
obtained from CTRW numerical convolution (symbols), and theoretical predictions; Eq. (\ref{eq_M2:P(x,t)_LDT_Final}) in dashed lines, Eq. (\ref{eq_M2:Z(q)_and_P(x,t)_asymp}) in solid lines,  and CLT in orange lines. The waiting times are drawn from the Erlang distribution (shape parameter = rate = $2$), with $\beta=1.5,3$ corresponding to triangles and circles.
Here, $t/\langle\tau\rangle=1$, as we focus on the scale where a particle jumps on average only once.
While the full Lambert solution fits well with the data for finite $|x|$, the logarithmic approximation in Eq. (\ref{eq_M2:Z(q)_and_P(x,t)_asymp}) does not.
}\label{fig_P(x,t)_Laplace_Erlang_mismatch}
\end{figure}

\section{Convergence of the two ARFS}\label{ap:Convergence_sokolov_to_us}
\begin{figure}[ht]
\centering
\includegraphics[width=0.5\textwidth]{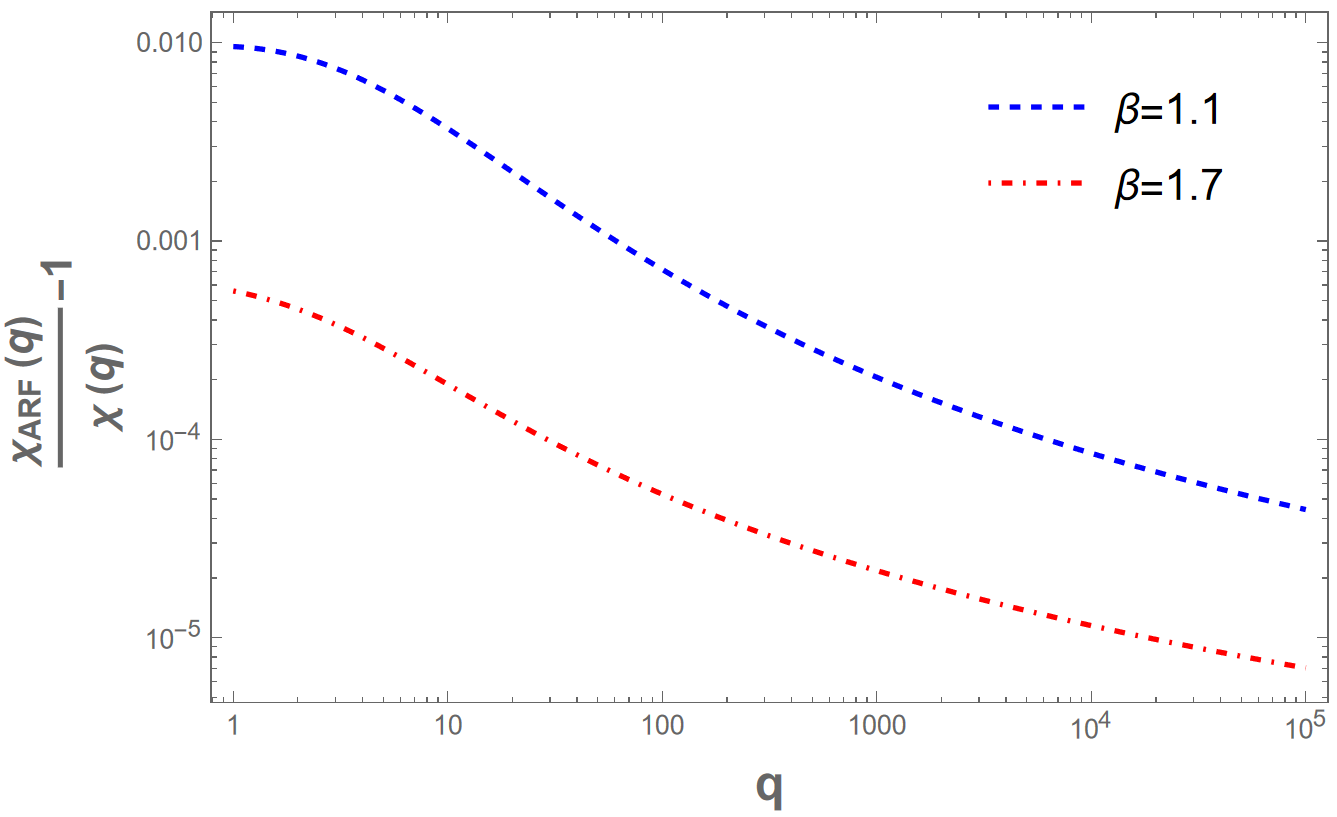}
\caption{
Convergence of the rate function of $P(x,t)$ obtained by our method using ${\large\chi}_{\scalebox{0.55}{$\mathrm{ARF}$}}(q)$ from Eq. (\ref{eq_M2:z(q)}) to the one described by the formalism of Ref. \cite{PhysRevE.103.042116}, detailed in Eq. (\ref{eq_M2:sokolov_RF_general_beta}). The parameters are: $A=5$, $C_A=1$, $C_{A+1}=-5$, with $\beta=1.1$ (blue), and $\beta=1.7$ (red).}
\end{figure}

\section{Leading scaling using the rate function formalism}\label{ap:leading_scaling_sokolov}
In this appendix, we derive the leading scaling behavior of the rate function using the rate function formalism. We substitute Eq. (\ref{eq_M2:sokolov_RF_zeta_star}) into Eq. (\ref{eq_M2:sokolov_RF_general_beta}) and then approximate the Lambert function using a logarithm. The resulting expression is taken in the limit $ q \to \infty $. Therefore, we do not recommend using it as a practical approximation.
Inserting Eq. (\ref{eq_M2:sokolov_RF_zeta_star}) into Eq. (\ref{eq_M2:sokolov_RF_general_beta})
\begin{eqnarray}
    \chi(q)\approx&-&\frac{C_{A+1}}{C_A}-\zeta_0^{1/\beta}q
    \left\{\frac{\alpha^\beta}{\zeta_0}W_0^{1-1/\beta}+\frac{\beta-2 A -4}{2} \frac{\log (W_0^{1/\beta })}{W_0^{1/\beta }} +(A+1) \frac{\log (q)}{W_0^{1/\beta }}+\frac{\tilde{r}_0}{W_0^{1/\beta }}\right\},   
\end{eqnarray}
with $\tilde{r}_0 = r_0 - \frac{\beta-2 A -4}{2}\log(\zeta_0^{1/\beta })$. We neglect $\tilde{r}_0 $, and approximate the Lambert as a log
\begin{eqnarray}
    \chi(q)\approx&-&\frac{C_{A+1}}{C_A}-\zeta_0^{1/\beta}q
    \left\{\frac{\alpha^\beta}{\zeta_0}\log\left(\zeta_0 \zeta_1z^{B_5}\right)^{1-1/\beta }+\frac{\beta-2 A -4}{2} \frac{\log (\log\left(\zeta_0 \zeta_1z^{B_5}\right)^{1/\beta })}{\log\left(\zeta_0 \zeta_1z^{B_5}\right)^{1/\beta }} +(A+1) \frac{\log (q)}{\log\left(\zeta_0 \zeta_1z^{B_5}\right)^{1/\beta }}\right\}
    \nonumber\\
\end{eqnarray}
Neglecting the middle term, which slowly decays as $q \to\infty$, and approximating $\log\left(\zeta_0 \zeta_1q^{B_5}\right) \approx \log\left(q^{B_5}\right)$, gives 
\begin{eqnarray}
    \chi(q)&\approx&
    -\frac{C_{A+1}}{C_A}-\zeta_0^{1/\beta}q
    \left\{\frac{\alpha^\beta}{\zeta_0}B_5^{1-1/\beta}\log\left(q\right)^{1-1/\beta } +B_5^{-1/\beta}(A+1) \log (q)^{1-\frac{1}{\beta}} \right\}
    \nonumber\\
    &=&-\frac{C_{A+1}}{C_A}-\zeta_0^{1/\beta}
    \left\{\frac{\alpha^\beta}{\zeta_0}B_5^{1-1/\beta} +B_5^{-1/\beta}(A+1)  \right\}q\log (q)^{1-\frac{1}{\beta}}
\end{eqnarray}
Namely, we got that in the limit of $q\to\infty$, $\chi(q)\sim -\frac{C_{A+1}}{C_A}-const\cdot q\log (q)^{1-\frac{1}{\beta}}$.

%
%
%
%
%
%
%
%
%
%
%
\bibliographystyle{apsrev4-2}
\bibliography{mybib}

\end{document}